%% file: MAIN.tex
\documentclass[final,1p]{elsarticle}

\usepackage{natbib}
\usepackage{lineno}
\usepackage{hyperref}
\usepackage{amsmath,amsfonts,amssymb,amsthm}
\usepackage{graphicx} 

\usepackage[english]{babel}

\biboptions{longnamesfirst,square,semicolon}

\newcommand{\eps}{\varepsilon}	        
\newcommand{\vect}[1]{\mathbf{#1}}	
\newcommand{\Par}[1]{\left({#1}\right)}	
\newcommand{\LTE}{\mathcal{E}}		
\newcommand{\dx}{\Delta x}     

\journal{Journal of Computational Physics}

\begin{document}

\renewcommand*{\today}{December 20, 2011}

\begin{frontmatter}

\title{A high order cell-centered semi-Lagrangian scheme for multi-dimensional kinetic simulations of neutral gas flows}

\author[yg]{Y. G\"{u}\c{c}l\"{u} \corref{cor}}
\ead{yguclu@msu.edu}

\author{W. N. G. Hitchon}
\ead{hitchon@engr.wisc.edu}

\address{Department of Electrical and Computer Engineering, University of Wisconsin, 1415 Engineering Dr., Madison, WI 53706, USA}

\input{Abstract.tex}


\cortext[cor]{Corresponding author.}
\fntext[yg]{Present address: Department of Mathematics, Michigan State University, East Lansing, MI 48824, USA.}

\end{frontmatter}



\input{Introduction.tex}

\input{CS_RemappingRule.tex}
\input{NumericalErrorsInASemiLagrangianScheme.tex}
\input{ModifiedEquationAnalysis.tex}
\input{HighOrderScheme.tex}
\input{3DAnalysis.tex}
\input{MonotonicVariations.tex}
\input{NumericalTests.tex}
\input{Conclusions.tex}
\input{Acknowledgments.tex}
\input{Appendix.tex}

\nocite{*}
\bibliographystyle{model1-num-names}
\bibliography{GucluHitchon2011}

\end{document}

%% file: Abstract.tex
\begin{abstract}

The term `Convected Scheme' (CS) refers to a family of algorithms, most usually applied to the solution of Boltzmann's equation, which uses a method of characteristics in an integral form to project an initial cell forward to a group of final cells.
As such the CS is a `forward-trajectory' semi-Lagrangian scheme.
For multi-dimensional simulations of neutral gas flows, the cell-centered version of this semi-Lagrangian (CCSL) scheme has advantages over other options due to its implementation simplicity, low memory requirements, and easier treatment of boundary conditions.
The main drawback of the CCSL-CS to date has been its high numerical diffusion in physical space, because of the 2$^{\text{nd}}$ order remapping that takes place at the end of each time step.
By means of a Modified Equation Analysis, it is shown that a high order estimate of the remapping error can be obtained a priori, and a small correction to the final position of the cells can be applied upon remapping, in order to achieve full compensation of this error.
The resulting scheme is 4$^{\text{th}}$ order accurate in space while retaining the desirable properties of the CS: it is conservative and positivity-preserving, and the overall algorithm complexity is not appreciably increased.
Two monotone (i.e. non-oscillating) versions of the fourth order CCSL-CS are also presented: one uses a common flux-limiter approach; the other uses a non-polynomial reconstruction to evaluate the derivatives of the density function.
The method is illustrated in simple one- and two-dimensional examples, and a fully 3D solution of the Boltzmann equation describing expansion of a gas into vacuum through a cylindrical tube.

\end{abstract}

%% file: Introduction.tex
\section{Introduction}

In this paper we describe a straightforward and easily implemented scheme for solution of the Boltzmann equation (or similar kinetic equations) with a number of desirable properties (straightforward conservation, positivity preservation and simplicity of implementation). The starting point is a solution of the kinetic equation in integral form, which uses the method of characteristics to take an initial cell and propagate it forward in time until it is remapped onto the phase space mesh. We refer to the scheme as a `Convected Scheme' (CS) to reflect the fact that it focusses on describing the convective motion of the fluid \cite{Hitchon1999}. More recently a terminology has arisen that would call this a `forward semi-Lagrangian' scheme.

A version of this scheme follows initial cells until they are almost totally depleted by collisions \cite{Feng2000,Christlieb2000}. In this `Long Lived Moving Cell' (LLMC) scheme, collisions remove particles from the cell periodically and the scheme `remaps' those particles to other locations in velocity space.
The earlier versions of the scheme limited the duration of the ballistic move to a small fraction of a collision time, after which the entire contents of the cell were remapped \cite{Hitchon1999}, the remapping reflecting the effects of the ballistic move (i.e. employing the method of characteristics) and for a minority of the population, reflecting in addition the effects of collisions.

The benefit of moving the initial cell is essentially that it allows us to focus on the conservation of various quantities associated with a single initial cell (number, energy, momentum) when the contents of that cell are remapped to the phase space mesh.
Particular attention was paid in plasma simulations to handling the ballistic motion near the points at which particles bounce, and in the vicinity of the zero energy cell, which turns out to be a little difficult to handle while conserving energy \cite{Parker1993b,Parker1994}.

In the earlier work, the order of the scheme was not considered to be of great concern, since the methods were integral methods and the number of remappings per collision time (which is often the critical physical time period) or even during the entire simulation was small. The remapping errors were thus believed to be modest. Nevertheless, if the desirable properties of the scheme can be retained in a higher order version, this would represent an improvement. 

It is the purpose of this paper to report a simple set of modifications (essentially a small correction to the final position of the MC upon remapping) which changes the order of the spatial move from second to fourth. 
The basic CS is unchanged by these modifications, in that conservation laws can be built into the scheme as before; the ease of implementation is similarly basically unaffected.
(While a similar procedure can certainly be applied to the energy/velocity move, its effects on energy conservation mean that its desirability is less clear cut. In this work, we focus on neutral particle kinetics, where the velocity is constant.)


Section \ref{sec:CS_RemapRule} briefly reviews the Convected Scheme, with particular emphasis on the ballistic operator and the remapping rule it employs.
Section \ref{sec:NumericalErrorsAndMotivation} discusses the qualitative and quantitative behavior of the numerical errors arising in the semi-Lagrangian solution of a transport problem, and it gives the motivations for the development of a high-order ballistic operator.
In Section \ref{sec:MEA}, a closed-form solution for the leading error of the Cell-Centered Semi-Lagrangian CS is obtained, which is amenable to be directly compensated by the use of `small corrections' to the final position of the MCs, as described in Section \ref{sec:HighOrderBallisticOperator}.
When a uniform velocity-field is considered, the 1D results generalize naturally to a 2D or 3D domain, by simply applying the corrections independently for each of the Cartesian axes.
Section \ref{sec:3DAnalysis} presents a simplified treatment for the solution of the 3D continuity equation with non-uniform velocity field.
Section \ref{sec:MonotonicOption} describes two different strategies for suppressing the oscillating behavior in the numerical solution which arises when the initial conditions are not sufficiently smooth (for smoothness, a continuous second order derivative would be required): the resulting scheme correctly preserves the monotonicity of the solution for generic initial conditions.
Section \ref{sec:NumericalTests} describes applications to solution of the 1D advection equation with uniform velocity, 2D continuity equation with non-uniform velocity, and 3D-3V
Boltzmann equation describing the gas expansion from a nozzle through a tube into vacuum. This 3D kinetic test was chosen for its applicability to space thruster simulations.

%% file: CS_RemappingRule.tex
\section{The Convected Scheme remapping rule}
\label{sec:CS_RemapRule}

The Convected Scheme (CS) is a mesh-based method for the solution of Boltzmann's equation, which describes the time evolution of the distribution function $f\Par{t, \vect{x}, \vect{v}}$ of a set of identical particles. The scalar field $f\Par{t, \vect{x}, \vect{v}}$ can be interpreted as the particle number density in phase space and it has dimensions of $[m^{-6}s^3]$. The differential form of Boltzmann's equation is
\begin{equation} \label{eq:Boltzmann}
	\frac{\partial f}{\partial t} + \vect{v}\cdot\frac{\partial f}{\partial\vect{x}} + \frac{\vect{F}}{m}\cdot\frac{\partial f}{\partial\vect{v}}
	= \left[\frac{\partial f}{\partial t}\right]_{coll}
\end{equation}
where $\vect{F}\Par{t,\vect{x},\vect{v}}$ is the force field acting on each particle, and the term on the right-hand side is the \emph{Boltzmann collision integral}, which describes the effect of collisions among particles. Equation \eqref{eq:Boltzmann} is expressed in an Eulerian form, and 7 scalar independent variables are present. Equivalently, one can focus on a Lagrangian trajectory $\Par{\vect{x}(t),\vect{v}(t)}$ in phase space, defined as:
\begin{equation} \label{eq:characteristics}
\left\{\begin{split}
	\dfrac{d\vect{x}}{dt} &= \vect{v}\,(t) \\
	\dfrac{d\vect{v}}{dt} &= \frac{1}{m}\,\vect{F}\Par{t,\vect{x},\vect{v}} \\
\end{split}\right.
\end{equation}
with initial conditions $\Par{\vect{x}_0,\vect{v}_0}$ at $t=t_0$. Substituting \eqref{eq:characteristics} into \eqref{eq:Boltzmann}, the $l.h.s.$ can be identified as the total derivative of $f$, and \eqref{eq:Boltzmann} is rewritten as
\begin{equation} \label{eq:Boltzmann-Lagr}
	\frac{Df}{Dt} = \left[\frac{\partial f}{\partial t}\right]_{coll},
\end{equation}
which states that the time rate of change of the distribution function along the trajectory \eqref{eq:characteristics} is only determined by the collision operator. Hence, in the absence of collisions $f$ is constant along each phase-space trajectory. Following naturally from these considerations, the Convected Scheme solves \eqref{eq:Boltzmann-Lagr} using a time-splitting procedure:
\begin{subequations}\label{eq:TimeSplitting}
\begin{enumerate}
	\item first, the distribution function is evolved for one time step according to the collision operator only
	\begin{equation}\label{eq:CollisionStep}
		\frac{\partial f}{\partial t} = \left[\frac{\partial f}{\partial t}\right]_{coll}
	\end{equation}
	\item then, the updated distribution function is advected for one time step along the characteristic trajectories, according to the collisionless equation
	\begin{equation} \label{eq:NonCollisional}
		\frac{Df}{Dt} = 0\ .
	\end{equation}
\end{enumerate}
\end{subequations}
Whether to execute the steps above in this order or not is really an arbitrary choice, since it only depends on the definition of time-step. The collision operator invariably involves the flux of particles in velocity space, but it is usually totally local in physical space. As such, step 1 is naturally solved on the phase-space mesh. As concerns step 2, the Convected Scheme really uses the integral form of \eqref{eq:NonCollisional}, with a single phase-space cell as the control-volume at $t=t_0$:
\begin{equation} \label{eq:NonCollisional-Integr}
	\iint_{C(t)}f\Par{t,\vect{x},\vect{v}}\ d\vect{x}\,d\vect{v}\ =\
	\iint_{C_0}f\Par{t_0,\vect{x}_0,\vect{v}_0}\ d\vect{x}_0 d\vect{v}_0
\end{equation}
where the \emph{r.h.s.} is just the total number of particles contained in cell $C_0$ at instant $t_0$, and $C(t)$ is the control-volume that evolves from $C_0$ by means of the map defined by \eqref{eq:characteristics}. For obvious reasons, $C(t)$ is called a Moving Cell (MC). The meaning of \eqref{eq:NonCollisional-Integr} is clear: it is merely a statement of mass conservation for a MC, during the ballistic motion. But to understand the practical utility of \eqref{eq:NonCollisional-Integr}, a suitable discretization of the variables must be introduced. The total number of particles in the initial phase-space cell $C_0$ is
\begin{equation*}
	N_0 = f_{ij}\Par{t_0}\,\Gamma_{ij} = N_{ij}\Par{t_0}
\end{equation*}
where $f_{ij}(t)$ is the average value of the distribution function in the phase-space cell centered at point $(\vect{x}_i,\vect{v}_j)$, and $\Gamma_{ij}$ is the corresponding volume in phase-space of the cell. It is evident from the last expression that the Eulerian phase-space mesh is needed not only to solve the collisional step, but also to give proper initial conditions to the ballistic operator:
\begin{equation*}
	N_{ij}^{MC}(t)\ \equiv\ N_{ij}\Par{t_0} \hspace{1cm} t \geq t_0\ .
\end{equation*}
At the end of the time step, the particles in the MC are remapped back onto the Eulerian mesh, where the collision operator is applied next. Alternatively, it is possible to evaluate the fraction of particles in the MC that have undergone a collision, and this permits one to remap to the mesh only those particles. In such cases, one can let the MC move for more than one time step, until the number of particles contained is a tiny fraction of the initial number, or until the geometrical distortion of the MC will require remapping to preserve accuracy. The application of the aforementioned ideas led to the Long-Lived Moving Cell (LLMC) version of the CS \cite{Feng2000,Christlieb2000}.

In any case, the exact time evolution of the MC can be evaluated analytically only in a few situations, and important approximations must be introduced regarding the following points:
\begin{enumerate}
	\item parametrization of the profile of $f$ in the MC,
	\item parametrization of the shape of the MC in phase-space,
	\item definition of the Lagrangian trajectories that are to be traced for each MC.
\end{enumerate}
Evidently, the choices made about the points above must be compatible with each other.\\
As concerns point 1, while the MC is always assumed to have a uniform density profile in physical space, it is usually considered to be a Dirac delta in velocity space.\\
As concerns point 2, the MC is usually obtained from the Cartesian product of a convex polyhedron in physical space and a convex polyhedron in velocity space. This is always the case when dealing with more than one dimension in physical space. Moreover, the MC is usually not deformable in velocity space, but it can usually undergo expansion/contraction and distortion in physical space. \\
As concerns point 3, there are three ways of handling the motion of the cell, that we have 
employed: first, we can move the cell center and assume the cell shape does not 
change; second, we can move the centers of cell faces and assume that each face does 
not change its orientation; third, we can move vertices of cells. In any case, a single initial velocity, equal to the central velocity $\vect{v}_j$, is given to the trajectories.

At the end of the ballistic move, particles must be placed back on the phase-space mesh according to the Convective Scheme remapping rule; for a domain in $\mathbb{R}^m$, this consists in depositing the particles from the Moving Cell onto the overlapped fixed cells $C_i$, according to the m-dimensional integral over $C_i$:
\begin{equation}\label{eq:CSRemap}
	N_i\ +\!=\ \oint_{C_i} n_{MC}(\vect{x}) \ d \vect{x}\ =\ \oint_{C_i \cap MC} n_{MC}(\vect{x}) \ d \vect{x}
\end{equation}
where $n_{MC}(\vect{x})$ is the m-dimensional density in the Moving Cell (MC).
The symbol $+\!=$ is the \emph{increment assignment operator}, meaning that the quantity on the \emph{l.h.s.} is increased by the amount on the \emph{r.h.s.}; hence, \eqref{eq:CSRemap} denotes the contribution to $N_i$ due to a given MC.
Since we are assuming that $n_{MC}$ is constant in the MC, the above rule can be simplified to
\begin{equation}\label{eq:CSRemap_uniform}
\begin{split}
	N_i\ +\!&=\ n_{MC}\ \oint_{C_i \cap MC} d \vect{x}\ =\ n_{MC}\ V_{C_i \cap MC}\ = \\
	&=\ n_{MC}\ V_{MC}\ \frac{V_{C_i \cap MC}}{V_{MC}}\ =\ N_{MC}\ F_i 
\end{split}
\end{equation}
where the symbol $V$ is used to denote the volumes in $\mathbb{R}^m$, and $F_i$ is called the "overlapping fraction".
Obviously, the sum of all the overlapping fractions is one:
\begin{equation*}
	\sum_i {F_i}\ =\ \sum_i {\left(\frac{V_{C_i \cap MC}}{V_{MC}}\right)}\ =\ \frac{V_{MC}}{V_{MC}}\ =\ 1
\end{equation*}

In the simplest version of a Convected Scheme, the MCs are rigidly translated according to the motion of their center, and hence the density in each MC is uniform and constant during the time step. In the semi-Lagrangian version at hand, the MCs are remapped onto the mesh at the end of each time-step. This scheme will be referred to as the \emph{Cell-Centered Semi-Lagrangian Convected Scheme} (CCSL-CS).

As we shall see, the CCSL-CS has many desirable properties: it is conservative, it preserves positivity, and it has no phase-error. On the other hand, the accuracy of the scheme is only $2^{\text{nd}}$ order in space, the leading error being a strong numerical diffusion term which has a non-linear dependence on the Courant parameter. In order to increase the accuracy of the scheme, two different approaches can be used:
\begin{enumerate}
	\item allowing for a non-uniform density in the MC;
	\item given an a priori estimate of the remapping error, applying a small correction to the final position of the MC to compensate for it.
\end{enumerate}
The former option complicates significantly the evaluation of the integral in \eqref{eq:CSRemap}, since \eqref{eq:CSRemap_uniform} is not applicable. Seeking a more numerically efficient method, the latter option is investigated hereafter.

%% file: NumericalErrorsInASemiLagrangianScheme.tex
\section{Numerical errors in a semi-Lagrangian scheme}
\label{sec:NumericalErrorsAndMotivation}

In this section we discuss the errors which arise in a semi-Lagrangian (SL) solution of a transport problem.
We point out that the order of the errors in a SL scheme without a Courant limit
is different from the order of the same scheme if a Courant limit were assumed to hold.

The ballistic operator of the Convected Scheme is a discrete approximation of \eqref{eq:NonCollisional}, and it is effectively decoupled into two different phases:
\begin{enumerate}
	\item the integration of the characteristic trajectories forward in time;
	\item the remapping of the transported field from the Lagrangian mesh to the fixed (Eulerian) mesh.
\end{enumerate}
As such, the CS ballistic operator falls into the category of `forward-trajectory semi-Lagrangian methods'.
Semi-Lagrangian (SL) methods were first introduced by Robert at the beginning of the eighties \cite{Robert1981}, and extensively applied to weather forecasting, oceanography, and fluid dynamics in general.
Most early SL methods traced the characteristic trajectories backward in time, from time $t$ to $t-\Delta t$, so that the interpolation (or the remapping) was done from an Eulerian to a Lagrangian mesh, at time step $t-\Delta t$.
SL schemes that project information forward along the trajectory have been in existence for some time (see the earlier versions of the CS, for example) but the terminology `forward-trajectory' scheme seems to have originated with Purser in the early nineties \cite{Purser1991}.

As typical of SL schemes \cite{Staniforth1991}, there is no numerical stability requirement that defines a maximum allowed time step $\Delta t_{\text{max}}$ for a given mesh spacing $\Delta x$, i.e.\ no Courant-Friedrichs-Lewy (CFL) \cite{CFL1928} limit applies to this explicit scheme.
(Nevertheless, a `deformational CFL limit' may still apply \cite{Purser1991}, which is usually much less restrictive.)
As a consequence, provided that the integration of the characteristics matches its own stability requirements, the time discretization $\Delta t$ and the space discretization $\Delta x$ can be chosen independently, based on accuracy considerations.

To assess the accuracy of the scheme, there are two complementary approaches:
\begin{enumerate}
\item a-priori estimates of the Local Truncation Error (LTE), i.e. the error introduced by one time-step of the algorithm, as a function of $\Delta t$ and $\Delta x$;
\item a-posteriori analysis of the Global Error (GE), i.e. the error in the solution after a certain integration time T.
\end{enumerate}
If it can be shown that the $\text{LTE} \le \alpha \Delta t$ as $\Delta t,\Delta x \rightarrow 0$ (in which case the scheme is said to be \emph{consistent}), and provided the scheme is also \emph{stable}, then \emph{convergence} is obtained, i.e. the $\text{GE}(t=T) \rightarrow 0$ for $\Delta t,\Delta x \rightarrow 0$.
It is therefore natural to ask `how fast' the scheme \emph{converges} to the exact solution as $\Delta t,\Delta x \rightarrow 0$.

In a typical explicit Eulerian scheme the CFL limit requires $\Delta t = O\Par{\Delta x}$, which makes it easy to write the Local Truncation Error in a form like $\text{LTE} = O\Par{\Delta x}^s$.
For sufficiently smooth initial conditions, and `well-behaved' boundary-conditions, one can show that $\text{GE} = O\Par{\Delta x}^s$ as well, and for this reason $s$ is called the \emph{order of convergence} of the scheme.

The form of the LTE is more complicated for a SL scheme, especially if the trajectories are obtained with a high-order ODE integrator; here, no relation is introduced between $\Delta t$ and $\Delta x$, as they can be chosen independently.
To lowest order approximation, one can decompose the LTE into two separate contributions: if the trajectory integrator has order $r$, it will introduce an error $O\Par{\Delta t^{r+1}}$ at each time step; similarly, if the remapping algorithm has order $p$, it will introduce an error $O\Par{\Delta x^p}$ at each time step.
At time instant $t=T$ the Global Error in the solution will be approximately $N=T/\Delta t$ times the error at each time step, which yields an estimate for the rate of convergence of the form
\begin{equation*}
	\left\| f\Par{x,T} - f_{ex}\Par{x,T} \right\|\ =\
	A\,\Delta t^r + B\,\frac{\Delta x^p}{\Delta t} ,
\end{equation*}
where $f\Par{x,T}$ is the computed (numerical) solution at time $T$, and $f_{ex}\Par{x,T}$ is the \emph{exact} solution at the same time instant; a proper norm $\|\cdot\|$ should be chosen (usually $L_1$, $L_2$ or $L_\infty$) for the above relation to be meaningful.
The last result, although not obtained by a rigorous derivation, is fully consistent with some accurate analyses for a broad class of SL schemes \cite{Falcone1998}.

\begin{figure}[!ht]
\centering
\includegraphics[width=\textwidth]{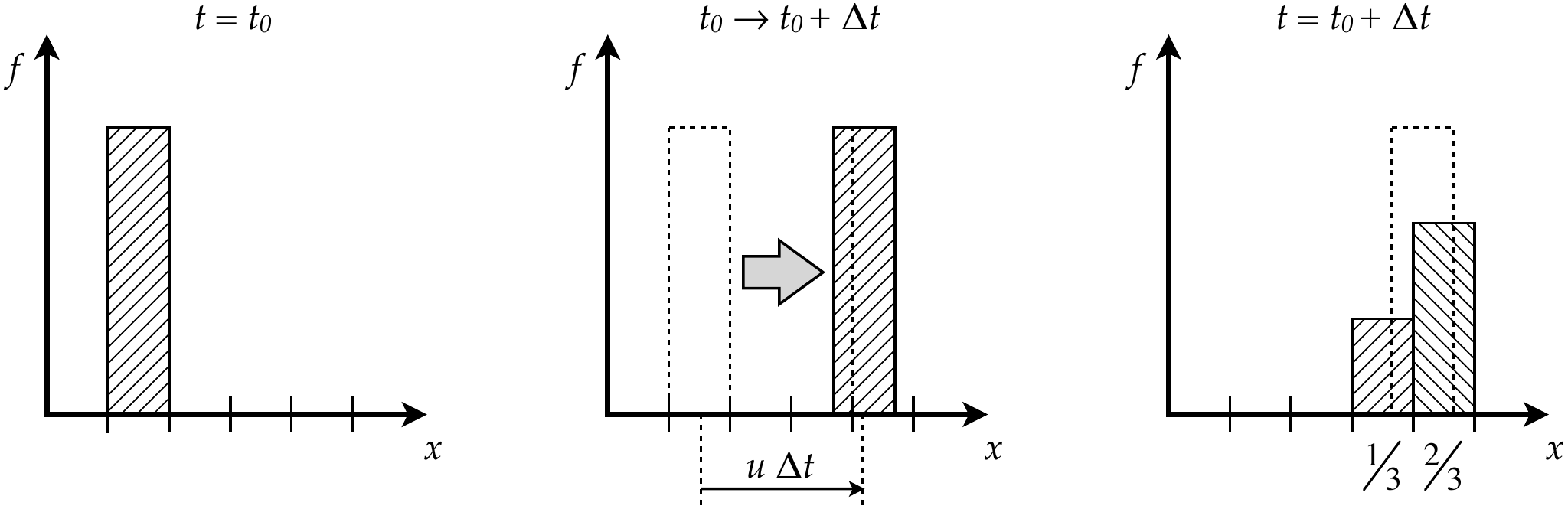}
\caption{
Ballistic operator of the 1D CCSL-CS, for a uniform mesh spacing.
The middle frame shows the Moving Cell (MC) after a time step.
Assuming that two thirds of the MC falls into the right hand cell which the MC overlaps, that is, the fraction of the MC's area which is in the right cell is two thirds, we would put two thirds of the particles in that cell.}
\label{fig:CCSL-CS.1D}
\end{figure}

For a kinetic simulation of a neutral gas, if the physical coordinates are fixed in an inertial reference frame and the gravitational force is neglected, then $\vect{F} = 0$ in Boltzmann's equation \eqref{eq:Boltzmann}, and the particles move in straight trajectories between two successive collisions.
In this situation, the Convected Scheme is a so-called `discrete velocity method' \cite{Broadwell1964}: at each instant in time and position in space, the velocity of the particles can only assume one of the prescribed values defined by the mesh in velocity space.
As a consequence, the velocity of the MC does not change during the ballistic phase, and no remapping is needed in velocity space; for each value $\vect{v}_\alpha$ of the velocity vector, the ballistic operator solves one passive advection equation with uniform constant velocity $\vect{v}_\alpha$ over the whole physical domain $\mathcal{D}\subset\mathbb{R}^3$:
\begin{equation*}
\frac{\partial}{\partial t} f\Par{\vect{x},\vect{v}_\alpha}
+ \vect{v}_\alpha\cdot\frac{\partial}{\partial\vect{x}}f\Par{\vect{x},\vect{v}_\alpha}
= 0, \qquad \vect{x}\in \mathcal{D}.
\end{equation*}
Figure \ref{fig:CCSL-CS.1D} shows the typical behavior of the ballistic operator of the CCSL-CS in a 1D case, with uniform mesh spacing $\Delta x$, for a single velocity $u_\alpha$.
In the very simple case at hand, a single MC is advected by a distance $u_\alpha \Delta t$, and it is then remapped on the 2 overlapped cells, the fraction being simply proportional to the overlap integral as given by \eqref{eq:CSRemap} (in 1D it is a length, in 2D an area, and in 3D a volume).
Already, the characteristic features of the scheme are clearly visible: positive attributes are
\begin{itemize}
\item since the advected quantity $f$ is inherently positive, and since the integrals are positive as well, the scheme is \emph{positivity preserving};
\item the centroid of the displaced density lies at its exact position, which means that the kinetic fluxes evaluated by the scheme are exact;
\end{itemize}
and negative attributes are:
\begin{itemize}
\item the density profile has flattened (i.e.\ two cells contain what was previously contained in one cell only) and one can easily imagine this process continuing until the density is uniform: \emph{numerical diffusion};
\item the density profile has become asymmetric: this is a higher-order error, the \emph{numerical dispersion}, which will be quickly masked by the numerical diffusion.
\end{itemize}
Intuitively, if the MC had moved by an exact multiple of $\Delta x$, its whole content would be remapped to one cell only, and there would be neither numerical diffusion nor numerical dispersion.
In the interesting case that the MC had moved by exactly $\Par{m+0.5}\Delta x$ ($m$ integer), we would have a situation with the maximum possible numerical diffusion (as the flattening would be maximum), but no numerical dispersion (since the final density profile would still be perfectly symmetric).
In general, the error depends on the fractional part $U_\alpha$ of the displacement, written as
\begin{equation*}
u_\alpha \Delta t = \Par{m_\alpha+U_\alpha} \Delta x,
\qquad m_\alpha \in \mathbb{I},
\quad U_\alpha \in \Par{-1,1}\subset \mathbb{R},
\end{equation*}
and of course, the error will depend on the density profile itself (e.g., there is no numerical error for a uniform density profile).

If the following notation is introduced:
\begin{eqnarray*}
	U_\alpha^+ = \frac{1}{2} \left( U_\alpha + |U_\alpha| \right)\ \geq 0 \\
	U_\alpha^- = \frac{1}{2} \left( U_\alpha - |U_\alpha| \right)\ \leq 0
\end{eqnarray*}
the 1D CCSL-CS algorithm can be recast as a simple finite-difference single-step scheme with a 3-point stencil:
\begin{equation*}
	f_i^{k+1}\ =\
	f_{i+m_\alpha-1}^k U_\alpha^+\ +\
	f_{i+m_\alpha}^k (1-|U_\alpha|)\ -\
	f_{i+m_\alpha+1}^k U_\alpha^-
\end{equation*}
where $i$ is the index over space, and $k$ is the index over time.
Apart from the integer shift of $m_\alpha$ cells, the last expression is equivalent to the 1D Donor-cell scheme; it is important to note that this equivalence only holds in the case of uniform velocity, while for a generic velocity field there is an important difference: \emph{the velocity in the CCSL-CS is not defined at the cell edges, but at the cell centers}.
It should be pointed out that the analogy between the CCSL-CS and the Donor-cell scheme only holds for 1D domains, as the two schemes generalize differently to higher dimensions.

Based on the aforementioned considerations, one can easily adapt the results for the 1D  Donor-cell scheme, obtained in \cite{Margolin1998} by a rigorous Modified Equation Analysis (MEA) in the hypothesis of `sufficiently smooth' density profile, to get a two-term expression for the \emph{Modified Equation} of the 1D CCSL-CS:
\begin{equation} \label{eq:1D_ModifiedEq}
\frac{\partial f}{\partial t} + u_\alpha\frac{\partial f}{\partial x}\ =\
\underbrace{\frac{\Delta x^2}{2\Delta t} \left( |U_\alpha| - U_\alpha^2 \right) \frac{\partial^2 f}{\partial x^2}}_\text{numerical diffusion}\ +\
\underbrace{\frac{\Delta x^3}{6\Delta t} \left( -U_\alpha + 3|U_\alpha|U_\alpha - 2U_\alpha^3  \right) \frac{\partial^3 f}{\partial x^3}}_\text{numerical dispersion} + H.O.T.\ ,
\end{equation}
where $H.O.T.$ stands for `higher order terms'.
The first term on the right hand side ($r.h.s.$) of the modified equation \eqref{eq:1D_ModifiedEq} is the leading error of the CCSL-CS, which scales with $\Delta x^2/\Delta t$.
Such an error term is linearly proportional to the second spatial derivative of $f(x)$, and it has the form of a Numerical Diffusion:
\begin{equation*}
	\mathcal{E}_1 = \frac{\partial}{\partial x}\left(D_\alpha\frac{\partial f}{\partial x}\right) \hspace{1cm}
	D_\alpha = \frac{\Delta x^2}{2\Delta t} \left( |U_\alpha| - U_\alpha^2 \right)
\end{equation*}
Such a numerical diffusion error, which may seem relatively innocuous, is actually quite pernicious to the numerical results, not only for the magnitude of the error, but also for the non-uniform distribution of the error in phase-space.
Hence, it is worth analyzing the behavior of this term in some detail.

For a kinetic gas simulation, the velocities $u_\alpha$ are defined by the velocity mesh, and the time-step $\Delta t$ is chosen based on the accuracy of the time-splitting procedure in \eqref{eq:TimeSplitting} (a reasonable value of $\Delta t$ is between $1/10$ and $1/5$ of the average collision time); as a consequence, $\Delta x$ remains the only free parameter that influences the numerical diffusion coefficient, through the factor $\Delta x^2$ and the parameter $U_\alpha$.

\begin{figure}[!ht]
\centering
\includegraphics[width=0.75\textwidth]{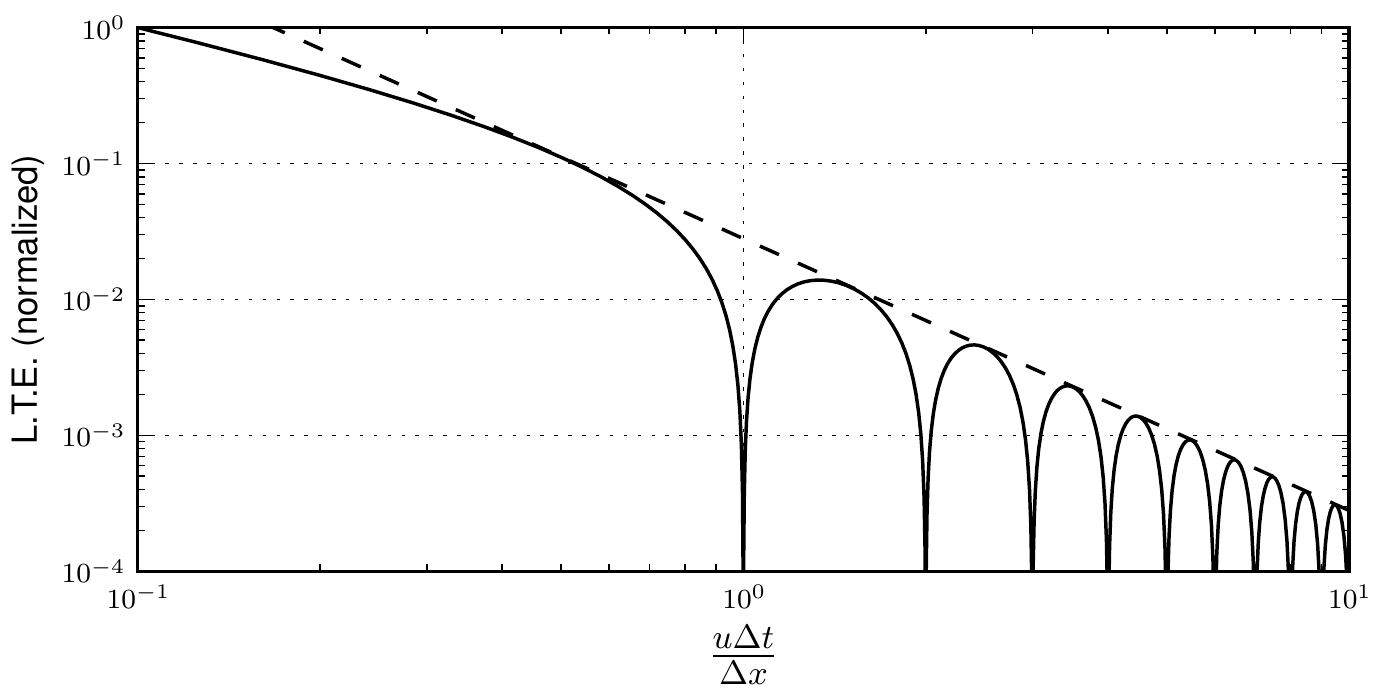}
\caption{
Numerical diffusion of the CCSL-CS in the solution of the 1D advection equation with uniform velocity $u$.
The Local Truncation Error (normalized to 1) is plotted against the `Courant parameter' $C_p = u\Delta t/\Delta x$ (solid line).
The graph is obtained by keeping $u\Delta t$ constant (as typical in a kinetic simulation of a neutral gas), and reducing progressively the mesh-size $\Delta x$.
The reference dashed line is multiply tangent to the curve of the LTE, and decreases as $\Delta x^2$.
}
\label{fig:Error_vs_Cp}
\end{figure}

Figure \ref{fig:Error_vs_Cp} shows the non-monotonic decrease of the leading numerical diffusion error for decreasing values of the mesh spacing $\Delta x$, for a fixed velocity $u_\alpha$ and fixed time-step size $\Delta t$.
The information provided by this picture is that the decrease of $\Delta x$ reduces the error on average, but may in fact increase the error for certain velocities $u_\alpha$ which were positioned on a monotonically increasing region of the curve.
The non-linear relation between the diffusion error and the velocity $u_\alpha$ produces strong anisotropy in the diffusion term when 2D and 3D spatial domains are considered.
For example, the numerical diffusion error in 2D is
\begin{equation}\label{eq:NumericalDiffusion2D}
	\mathcal{E}_1^{\text{(2D)}} =
	D_x \frac{\partial^2 f}{\partial x^2} +
	D_y \frac{\partial^2 f}{\partial y^2},
	\qquad \text{with} \quad
	\begin{cases}
	D_x = \frac{\Delta x^2}{2\Delta t} \Par{|U_\alpha| - U_\alpha^2} \\
	D_y = \frac{\Delta y^2}{2\Delta t} \Par{|V_\alpha| - V_\alpha^2}
	\end{cases}
\end{equation}
where $U_\alpha$ and $V_\alpha$ are obtained from the fractional part of the displacements $u_\alpha \Delta t$ and $v_\alpha \Delta t$, along the $x$ and $y$ directions respectively.

\begin{figure}[!ht]
\includegraphics[width=\textwidth]{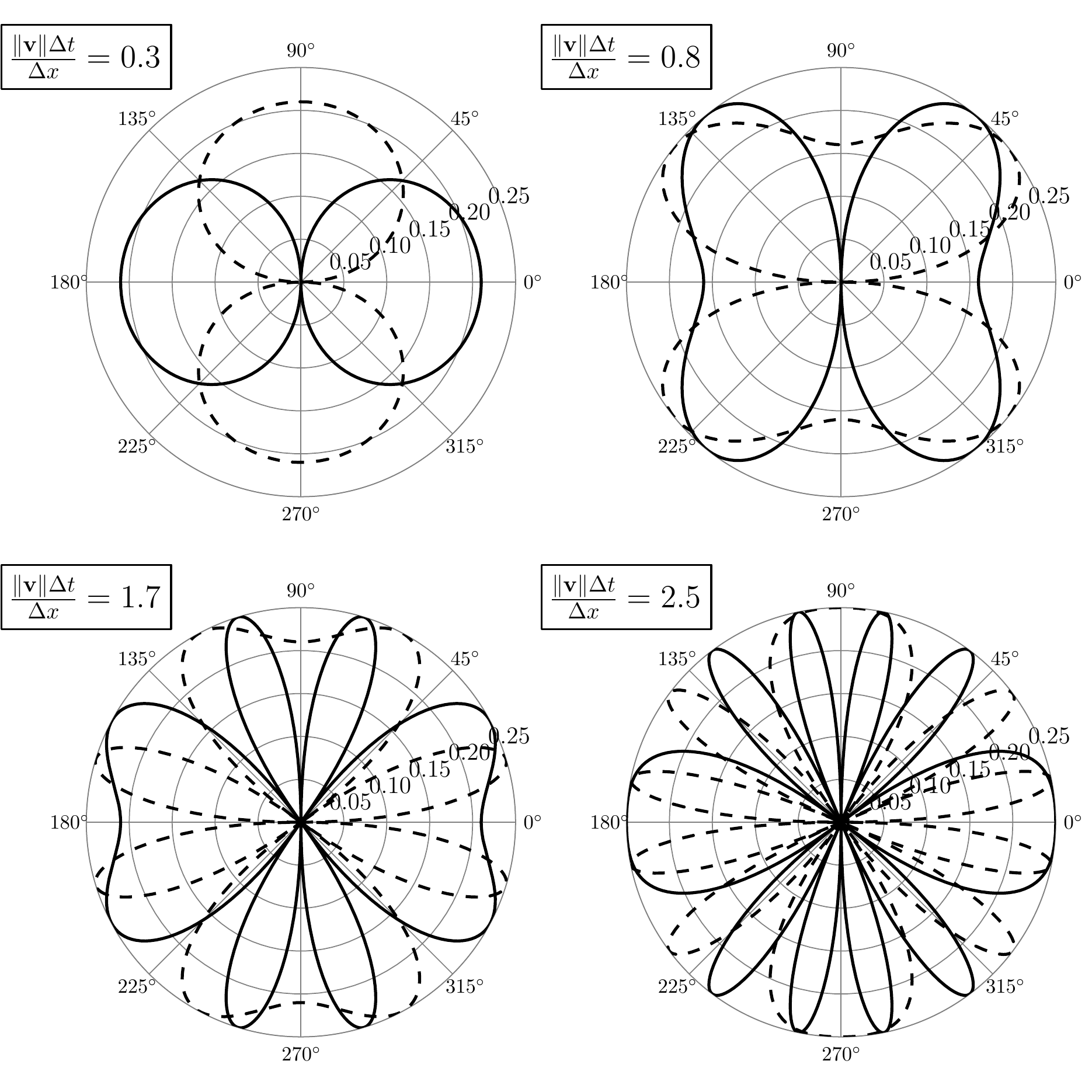}
\caption{
Numerical diffusion of the CCSL-CS scheme in the solution of the 2D advection equation with uniform velocity.
For four different values of the speed $\|\vect{v}\|$, the diffusion coefficients along the directions $x$ and $y$ are plotted as functions of the angle $\theta = \vect{v}\cdot \hat{\vect{x}}/\|\vect{v}\|$: the solid lines represent $D_x\Par{\theta}$, while the dashed lines represent $D_y\Par{\theta}$.
The value of a diffusion coefficient is obtained from the local radius $r\Par{\theta} \in [0,0.25]$ of the corresponding curve in the polar plot, multiplied by $\Delta x^2/\Delta t$ (see \eqref{eq:NumericalDiffusion2D}).
}
\label{fig:DxDy_vs_Angle}
\end{figure}

Assuming $\Delta x = \Delta y$, the diffusion coefficients in the two directions are most likely to be widely different from each other, i.e.\ \emph{the numerical diffusion of the 2D CCSL-CS is strongly anisotropic}.
To better appreciate this effect, one can fix the magnitude $\|\vect{v}\| = S$ of the velocity vector, and write $u = S\cos\theta$ and $v = S\cos\theta$, to obtain $D_x\Par{\theta}$ and $D_y\Par{\theta}$; here $\theta = \vect{v}\cdot \hat{\vect{x}}/\|\vect{v}\|$ is the angle between the velocity vector and the $X$ axis of the grid.
Figure \ref{fig:DxDy_vs_Angle} presents polar plots of $D_x\Par{\theta}$ (solid line) and $D_y\Par{\theta}$ (dashed line), for a few different values of the velocity magnitude: the numerical diffusion is isotropic only for those values of $\theta$ where the two curves cross each other; for all other values, there is a preferential direction for diffusion.

Intuitively, in a 3D geometry the situation gets even worse, as the numerical diffusion has more degrees of freedom for causing distortion in the numerical solution.
As there is no way to avoid such a non-uniform distribution of the numerical error in phase-space, the problem can be mitigated only by globally reducing the magnitude of the error, a result that can be achieved by decreasing the mesh-size $\Delta x$.

In 3D, the solution converges with the same asymptotic behavior as in 1D and 2D, i.e.\ $\|\mathcal{E}\| = O\Par{\Delta x^2}$, but the number $N$ of cells in the domain increases as $\Delta x^{-3}$.
Hence, $N \propto \|\mathcal{E}\|^{-3/2}$, which means that to reduce the error by a factor of 2, the number of cells in the domain must be increased by $\approx 2.83$ times; accordingly, the memory requirement and the simulation time will increase by (at least) the same factor.
$N \propto \|\mathcal{E}\|^{-3/2}$ is a quite unfavorable scaling for a multi-dimensional kinetic simulation, as the computational limitations may easily prevent us from reaching the required accuracy in the solution.

The aforementioned considerations are the main reasons that motivated the development of a higher-order ballistic operator, which will be proved to be 4$^{\text{th}}$ order accurate in space, that is $\|\mathcal{E}\| = O\Par{\Delta x^4}$.
The error of the new scheme still behaves `erratically', similarly to the original scheme, but it is already much smaller in magnitude when a coarse mesh spacing is employed, and it has a much better scaling when $\Delta x$ is reduced.
In fact, the high-order scheme has $N \propto \|\mathcal{E}\|^{-3/4}$, so that to reduce the error by a factor of 2, the number of cells in the domain must be increased by $\approx 1.68$ times only.

The novelty of the implementation is in the fact that, even though it is the remapping stage which causes the numerical error, this stage is left unaltered; instead, the remapping error is compensated a-priori by applying `small corrections' to the final position of the MC prior to remapping.
As another way to put it, an error is deliberately introduced in the trajectory integrator (which was exact, since the particles move in straight trajectories), such that it \emph{exactly compensates for the diffusion and dispersion errors}, at least for sufficiently smooth density functions.

The resulting ballistic operator maintains the most desirable properties of the original CCSL-CS, (which are inherited from the fact that the remapping method is left unchanged): exact mass conservation, preservation of positivity, and simple and efficient implementation.
Moreover, the resulting scheme can be slightly modified to ensure monotonicity preservation: two different strategies will be presented to deal with this.

In the next Section, the LTE of the scheme is assessed when a non-uniform velocity field is considered.
This is necessary in order to account for the small modifications to the final position of the MC: in fact, once divided by the constant time-step $\Delta t$, a modification to the position can be equivalently regarded as a modification to the velocity.
Such corrections to the velocity field may be interpreted as `anti-diffusive' and `anti-dispersive' velocities: in this sense, our approach here is similar to the one used in the MPDATA algorithm \cite{Smolarkiewicz1984,Smolarkiewicz1990}, which is a high-order method based on the Donor-cell scheme.
The main advantage of the high-order version of the CCSL-CS over MPDATA is that the corrections are evaluated a-priori and in a single step, without the need for a multi-pass slowly converging procedure.

As an aside, the general treatment in the next Section, which does not require the velocity field to be uniform, opens the CCSL-CS to a range of applications it was not originally designed for.

%% file: ModifiedEquationAnalysis.tex
\section{Modified equation analysis for the `low-order' CCSL-CS}
\label{sec:MEA}

This section aims at the assessment of the error introduced by the 1D CCSL-CS, by means of a modified equation analysis (MEA) \cite{Warming1974}.
This is a standard procedure commonly applied to finite-difference schemes, based on a Taylor expansion of the solution around the point of interest $(t,x)$; as such, we will assume the transported function to be `sufficiently smooth'.

In order to make the following treatment more manageable in the presence of a non-uniform velocity field, it will be assumed that the Courant parameter $U=u\Delta t/\Delta x$ is less than 1.
As a consequence, $\Delta t = O\Par{\Delta x}$ and the order of accuracy in space is equivalent to the order of accuracy in time; moreover, the schemes effectively `lose' one order of accuracy: the original CCSL-CS will be said to be `1$^{\text{st}}$ order accurate' (although in fact, since the time step is not constrained by the Courant criterion, it is actually second order accurate in space), and the high-order CCSL-CS will be regarded as being `3$^{\text{rd}}$ order accurate' (again, the spatial order of accuracy is in fact fourth, since the time step is not in reality constrained to be small).

As a starting point, the equation to be solved is now the 1D continuity equation
\begin{equation}\label{eq:Continuity1D}
	\partial_t n + \partial_x \left(u\,n\right) = 0
\end{equation}
where the velocity $u(x,t)$ is not constrained to be constant in time or uniform in space.
Nevertheless, for the sake of clarity, it will be assumed that $u > 0$ unless otherwise stated.
In this situation, the 1D CCSL-CS is
\begin{equation}\label{eq:CS_1D}
	n_i^{k+1}\ =\ U_{i-1}^k\, n_{i-1}^k + \left(1-U_i^k\right)n_i^k
\end{equation}
which is, after substitution of the $3^{rd}$ order Taylor expansions about the point $\left(x_i,t^k\right)$,
\begin{equation} \label{eq:CS1D_expanded}
\begin{split}
	&\left(\Delta t\,\partial_t +\frac{\Delta t^2}{2}\partial_{tt}^2+ \frac{\Delta t^3}{6}\partial_{ttt}^3 \right) n + O\left(\Delta t^4\right) = \\
	-&\left(\Delta x\,\partial_x - \frac{\Delta x^2}{2}\partial_{xx}^2
	+\frac{\Delta x^3}{6}\partial_{xxx}^3\right) \left(U n\right)+ O\left(\Delta x^4\right)
\end{split}
\end{equation}
where $n \equiv n_i^k$ and $U \equiv U_i^k$.
As a first step, the time derivatives of $n$ are eliminated by means of the substitution:
\begin{equation} \label{eq:n_t}
	n_t\,\Delta t = -\left(U n\right)_x\Delta x + \LTE\,\Delta t
\end{equation}
where $\LTE$ is a (1$^{st}$ order) spurious term introduced by the numerical scheme.
Henceforth $\LTE$ will be referred to as the `modified equation error' (MEE), in order to distinguish it from the `local truncation error' (LTE), which was defined in Section \ref{sec:NumericalErrorsAndMotivation} as the error \emph{in the solution} introduced by the scheme after a single time-step.
When the solution is `sufficiently smooth', the LTE can be obtained from the integral of $\LTE$ over the time step, so that LTE $\approx \LTE \Delta t$.

An expression for $n_{tt}$ can be obtained by differentiating \eqref{eq:n_t} with respect to time, multiplying by $\Delta t$, and substituting the \emph{r.h.s.} of \eqref{eq:n_t} wherever the $n_t$ term appears:
\begin{equation} \label{eq:n_tt}
	n_{tt}\,\Delta t^2 =\ U\Par{Un}_{xx} \Delta x^2 + U_x\left(U n\right)_x \Delta x^2
	- \Par{U_t\,n}_x \Delta x\,\Delta t - \Par{U\,\LTE}_x \Delta x\,\Delta t + \LTE_t\,\Delta t^2
\end{equation}
where the last two terms on the \emph{r.h.s} are third order since they contain the MEE itself.
The same procedure can be repeated to get an expression for the third time derivative:
\begin{equation} \label{eq:n_ttt}
\begin{split}
	n_{ttt}\,\Delta t^3 =& -U^2\left(Un\right)_{xxx}\Delta x^3
	+ \left(2U_t\frac{\Delta t}{\Delta x} - 3UU_x\right)\left(Un\right)_{xx}\Delta x^3 \\
	&+ \left(2U_{xt}\frac{\Delta t}{\Delta x} - U_x^2 - UU_{xx}\right)\left(Un\right)_x\Delta x^3\\
	&+ U\left(U_t\,n\right)_{xx}\Delta x^2\Delta t + U_x\left(U_t\,n\right)_x\Delta x^2\Delta t
	- \left(U_{tt}\,n\right)_x\Delta x\,\Delta t^2 + H.O.T.
\end{split}
\end{equation}
where the $H.O.T.$ come from the terms containing the MEE, which are now fourth order.
Substituting \eqref{eq:n_t}, \eqref{eq:n_tt} and \eqref{eq:n_ttt} into \eqref{eq:CS1D_expanded}, one gets:
\begin{equation} \label{eq:ImplicitModifiedEqn}
\begin{split}
	n_t + \Par{un}_x\ &=\
	\frac{\Delta x^2}{2\,\Delta t}\,\Biggl[\Par{1-U}\Par{Un}_{xx} - U_x\Par{Un}_x + \Par{U_t\,n}_x \frac{\Delta t}{\Delta x}\Biggr] \\
	&+\ \frac{\Delta x^3}{6\,\Delta t}\,\Biggl[\Par{U^2-1}\Par{Un}_{xxx}+\Par{3UU_x-2U_t\frac{\Delta t}{\Delta x}}\Par{Un}_{xx} \\
	&\quad\quad+\Par{U_x^2+2U_{xx}-2U_{xt}\frac{\Delta t}{\Delta x}}\Par{Un}_x - U\Par{U_t\,n}_{xx}\frac{\Delta t}{\Delta x} \\
	&\quad\quad- U_x\Par{U_t\,n}_x\frac{\Delta t}{\Delta x} + \Par{U_{tt}\,n}_x\Par{\frac{\Delta t}{\Delta x}}^2\Biggr] \\
	&+\ \frac{1}{2}\,\Bigl[\Par{U\LTE}_x\Delta x - \LTE_t\,\Delta t \Bigr]\ + O\Par{\Delta x^3}
\end{split}
\end{equation}
where the \emph{r.h.s.} is the truncation error.
The fact that the expression for the MEE is formally implicit does not mean the problem is insoluble.
In fact, since the MEE is first order, the last term in brackets on the \emph{r.h.s.} is second order.
This implies that the leading-term (first order) of the MEE is already known at this stage, and it is just the first term on the \emph{r.h.s.} of \eqref{eq:ImplicitModifiedEqn}:
\begin{equation} \label{eq:LTE_1}
	\LTE_1 = \frac{\Delta x^2}{2\,\Delta t}\,\left[\Par{1-U}\Par{Un}_{xx} - U_x\Par{Un}_x + \Par{U_t\,n}_x \frac{\Delta t}{\Delta x}\right].
\end{equation}
The second order term of the MEE can be obtained by substituting $\LTE\approx\LTE_1$ into \eqref{eq:ImplicitModifiedEqn}.
Accordingly, the space and time derivatives of the MEE are approximated as
\begin{equation} \label{eq:LTE_x}
	\LTE_x \approx \frac{\Delta x^2}{2\,\Delta t}\,\left[\Par{1-U}\Par{Un}_{xxx} - 2U_x\Par{Un}_{xx}
	-U_{xx}\Par{Un}_x + \Par{U_t\,n}_{xx} \frac{\Delta t}{\Delta x}\right]
\end{equation}
and
\begin{equation} \label{eq:LTE_t}
\begin{split}
	\LTE_t \approx \frac{\Delta x^3}{2\,\Delta t^2}\,\Biggl[ &
	\Par{U^2-U}\Par{Un}_{xxx} + \Par{\Par{3U-2}U_x - 2U_t\frac{\Delta t}{\Delta x} }\Par{Un}_{xx} \\
	&+ \Par{\Par{U-1}U_{xx}+U_x^2 - 2U_{xt}\frac{\Delta t}{\Delta x} }\Par{Un}_x \\
	&+ \Par{1-U}\Par{U_t\,n}_{xx}\frac{\Delta t}{\Delta x} - U_x\Par{U_t\,n}_x\frac{\Delta t}{\Delta x}
	+ \Par{U_{tt}\,n}_x\Par{\frac{\Delta t}{\Delta x}}^2 \Biggr]
\end{split}
\end{equation}
where the approximation $n_t\Delta t\approx -\Par{Un}_x\Delta x$ was used instead of the full \eqref{eq:n_t}. The modified equation of the 1D CCSL-CS, subject to the hypothesis $u>0$, is obtained by substituting \eqref{eq:LTE_1} for $\LTE$, \eqref{eq:LTE_x} for $\LTE_x$, and \eqref{eq:LTE_t} for $\LTE_t$ into \eqref{eq:ImplicitModifiedEqn},
\begin{equation} \label{eq:ExplicitModifiedEqn}
\begin{split}
	n_t + \Par{un}_x\ &=\
	\frac{\Delta x^2}{2\,\Delta t}\,\Biggl[\Par{1-U}\Par{Un}_{xx} - U_x\Par{Un}_x + \Par{U_t\,n}_x \frac{\Delta t}{\Delta x}\Biggr] \\
	&+\ \frac{\Delta x^3}{6\,\Delta t}\,\Biggl[\Par{3U-2U^2-1}\Par{Un}_{xxx}+\Par{\frac{9}{2}U_x-6UU_x+U_t\frac{\Delta t}{\Delta x}}\Par{Un}_{xx} \\
	&\quad\quad +\Par{\frac{3}{2}U_{xx}-2UU_x-2U_x^2+U_{xt}\frac{\Delta t}{\Delta x}}\Par{Un}_x + \Par{2U-\frac{3}{2}}\Par{U_t\,n}_{xx}\frac{\Delta t}{\Delta x} \\
	&\quad\quad +2U_x\Par{U_t\,n}_x\frac{\Delta t}{\Delta x} -\frac{1}{2}\Par{U_{tt}\,n}_x\Par{\frac{\Delta t}{\Delta x}}^2\Biggr]\
	+\ O\Par{\Delta x^3}\ ,
\end{split}
\end{equation}
which shows that, for a given velocity field $u\Par{x,t}$, the PDE that is really solved by the scheme \eqref{eq:CS_1D} approximates the continuity equation $n_t + \Par{un}_x = 0$ only to first order.
In fact, \eqref{eq:ExplicitModifiedEqn} contains the spurious terms $\LTE_1 = O(\Delta x)$ and $\LTE_2 = O(\Delta x^2)$, which are the first two terms of the `modified equation error' (MEE).

%% file: HighOrderScheme.tex
\section{High-order version of the CCSL-CS}
\label{sec:HighOrderBallisticOperator}

In the following treatment a novel approach is used in order to devise a third order version of the CCSL-CS. From a Lagrangian view-point, the basic idea is to \emph{apply a correction to the position of the Moving-Cell before it is remapped}. Since a correction to the position is $\delta x = \delta u\, \Delta t$, one can think of a correction to the velocity as well. 

It is important to consider that the velocity field $u\Par{x,t}$ need not be the physical velocity field, which we call $u_0\Par{x,t}$ from now on. In fact, we can assume that
\begin{equation} \label{eq:U_expansion}
	u = u_0 + u_1 + u_2 \hspace{0.5cm} \text{or} \hspace{0.5cm}
	U = U_0 + U_1 + U_2
\end{equation}
where $U_1\propto\Delta x$ and $U_2\propto\Delta x^2$ will be evaluated by requiring \eqref{eq:ExplicitModifiedEqn} be a third order approximation to the `physical' continuity equation:
\begin{equation} \label{eq:continuity_approx}
	n_t + \Par{u_0n}_x = O\Par{\Delta x^3}.
\end{equation}
In other words, without changing the remapping rule on which the CS is based, a third order scheme can be obtained by applying first and second order corrections ($u_1$ and $u_2$) to the physical velocity field $u_0$. Such a method is direct (i.e.\ non iterative), and guarantees the compensation of both $\LTE_1$ and $\LTE_2$ for sufficiently smooth density profiles. After some algebra, the substitution of \eqref{eq:U_expansion} into \eqref{eq:ExplicitModifiedEqn} gives \eqref{eq:continuity_approx} only if
\begin{equation} \label{eq:U1}
	U_1\ =\ \frac{\Delta x}{2}\left[\Par{1-U_0}\frac{1}{n}\Par{U_0\,n}_x
	+ U_{0\,t}\,\frac{\Delta t}{\Delta x}\right],
\end{equation}
and
\begin{equation} \label{eq:U2}
\begin{split}
	U_2\ =\ \frac{\Delta x^2}{12}\ &\Biggl[
	\Par{1-3U_0+2U_0^2}\frac{1}{n}\Par{U_0\,n}_{xx}\ +\
	\Par{2U_0U_{0\,x}-3U_{0\,x}-4U_{0\,t}\frac{\Delta t}{\Delta x}}\frac{1}{n}\Par{U_0\,n}_x \\
	& + \Par{6-5U_0}\frac{1}{n}\Par{U_{0\,t}\,n}_x\frac{\Delta t}{\Delta x}\ +\
	2U_{0\,tt} \Par{\frac{\Delta t}{\Delta x}}^2 \Biggr].
\end{split}
\end{equation}
The expressions \eqref{eq:U1} and \eqref{eq:U2} formally contain the \emph{exact} space and time derivatives of $U_0$ and $n$, but for any practical situation, such derivatives will be evaluated numerically.
For \eqref{eq:continuity_approx} to hold, both $U_1$ and $U_2$ must be estimated with third order accuracy; accordingly, the finite difference schemes must be second order accurate in \eqref{eq:U1}, and first order in \eqref{eq:U2}.
Examples of the numerical evaluation of the derivatives are given in Section \ref{sec:NumericalTests}.

With the proper approximations for $U_1$ and $U_2$, the scheme that results from using the velocity corrections \eqref{eq:U_expansion} in the 1D CCSL-CS \eqref{eq:CS_1D} is third order accurate, conservative, and very simple to implement.
Most important, even if the high-order scheme does not ensure monotonicity preservation, it does preserve the positivity of the solution at all times, thanks to the fact that the remapping phase (inherently sign-preserving) is left unchanged.
If the initial conditions are `sufficiently smooth' (or, equivalently, the gradients are `well resolved'), the high-order scheme just formulated performs very well, as it also preserves the monotonicity in the solution.
When sharp gradients are present instead, spurious oscillations may arise in the solution: in such a situation, a `non-oscillatory' version of the present scheme is necessary, which will be discussed in Section \ref{sec:MonotonicOption}. 

When the velocity $u_{0}$ is constant and uniform, formulas \eqref{eq:U1} and \eqref{eq:U2} are further simplified: in such a situation, Cartesian 2D and 3D versions of the algorithm are obtained straightforwardly by applying independent corrections to each component of the velocity vector, resulting in a very compact and numerically efficient method.
In fact, all these properties make this approach particularly suitable for application to neutral gas kinetic simulations.

%% file: 3DAnalysis.tex
\section{3D analysis}\label{sec:3DAnalysis}

One might wonder whether the above procedure could be used to obtain a third order scheme for the advection equation in three dimensions, for a generic flow field. In fact, that would open the possibility to apply the CCSL-CS to the multi-dimensional solution of generic hyperbolic conservation equations (for example, the Euler equations that govern inviscid flow dynamics). Such an application is beyond the scope of this work, and it is opinion of the authors that a complete third order analysis in 3D can be carried out only with a considerable amount of algebra. Nevertheless, for completeness, a brief outline of the procedure to obtain a second-order scheme in 3D is given hereafter. \\
The equation to be solved is now the 3D continuity equation
\begin{equation*}
	\partial_t\, n + \partial_x \Par{u\,n} + \partial_y \Par{v\,n} + \partial_z \Par{w\,n} = 0,
\end{equation*}
where $u$, $v$, and $w$ are respectively the $x$, $y$, and $z$ components of the velocity vector. For the sake of clarity, it will be assumed that $u$, $v$, and $w$ are all $>0$, and that
\begin{equation*}
\begin{matrix}
	U = \dfrac{u\Delta t}{\Delta x} < 1, & &
	V = \dfrac{v\Delta t}{\Delta y} < 1, & &
	W = \dfrac{w\Delta t}{\Delta z} < 1 .
\end{matrix}
\end{equation*}
Under the hypothesis above, if a Cartesian grid with uniform spacing over each direction is considered, the 3D CCSL-CS can be written as a finite-difference scheme,
\begin{equation*}
\begin{split}
	n_{i,j,k}^{m+1}\ &=\ n_{i,j,k}^m\left(1-U_{i,j,k}^m\right)\left(1-V_{i,j,k}^m\right)\left(1-W_{i,j,k}^m\right) \\
	&+\ n_{i-1,j,k}^m\ U_{i-1,j,k}^m \left(1-V_{i-1,j,k}^m\right)\left(1-W_{i-1,j,k}^m\right) \\
	&+\ n_{i,j-1,k}^m\left(1-U_{i,j-1,k}^m\right) V_{i,j-1,k}^m \left(1-W_{i,j-1,k}^m\right) \\
	&+\ n_{i,j,k-1}^m\left(1-U_{i,j,k-1}^m\right)\left(1-V_{i,j,k-1}^m\right) W_{i,j,k-1}^m \\
	&+\ n_{i-1,j-1,k}^m\ U_{i-1,j-1,k}^m\ V_{i-1,j-1,k}^m \left(1-W_{i-1,j-1,k}^m\right) \\
	&+\ n_{i,j-1,k-1}^m\left(1-U_{i,j-1,k-1}^m\right) V_{i,j-1,k-1}^m\ W_{i,j-1,k-1}^m \\
	&+\ n_{i-1,j,k-1}^m\ U_{i-1,j,k-1}^m \left(1-V_{i-1,j,k-1}^m\right) W_{i-1,j,k-1}^m \\
	&+\ n_{i-1,j-1,k-1}^m\ U_{i-1,j-1,k-1}^m\ V_{i-1,j-1,k-1}^m\ W_{i-1,j-1,k-1}^m,
\end{split}
\end{equation*}
where $i$, $j$, and $k$ are the indices over the three spatial directions, and $m$ is the index over time.
A $2^{nd}$ order Taylor series expansion about the point $\left(x_i, y_j, z_k, t^m\right)$ gives:
\begin{equation} \label{eq:expansion_3D}
\begin{split}
	&n_t\,\Delta t + \Par{nU}_x\Delta x + \Par{nV}_y\Delta y + \Par{nW}_z\Delta z\ =\\
	&-n_{tt}\frac{\Delta t^2}{2} + O\Par{\Delta t^3} +
	\Par{nU}_{xx} \frac{\Delta x^2}{2} + \Par{nV}_{yy} \frac{\Delta y^2}{2} + \Par{nW}_{zz} \frac{\Delta z^2}{2} \\
	&+\Par{nUV}_{xy}\Delta x\,\Delta y + \Par{nVW}_{yz}\Delta y\,\Delta z + \Par{nWU}_{zx}\Delta z\,\Delta x
	  + O\Par{\Delta s^3},
\end{split}
\end{equation}
where $\Delta s = \sqrt{\Delta x^2+\Delta y^2+\Delta z^2}$.\\
As usual, we can get rid of $n_{tt}$ by differentiating the equality
\begin{equation}
	n_t\,\Delta t = -\Par{nU}_x\Delta x -\Par{nV}_y\Delta y -\Par{nW}_z\Delta z + \LTE\,\Delta t
\end{equation}
with respect to time. Seeking a second order scheme, the term containing the modified equation error $\LTE$ can be neglected, since it contributes to higher order terms only. After some algebraic manipulations, this gives:
\begin{equation} \label{eq:n_tt-3D}
\begin{split}
	n_{tt}\,\Delta t^2 &\simeq\
	     \Bigl[U\Par{nU}_x\Delta x + nVU_y\,\Delta y + nWU_z\,\Delta z - nU_t\,\Delta t\,\Bigr]_x\,\Delta x \\
	&+\ \Bigl[V\Par{nV}_y\Delta y + nWV_z\,\Delta z + nUV_x\,\Delta x - nV_t\,\Delta t\,\Bigr]_y\,\Delta y \\
	&+\ \Bigl[W\Par{nW}_z\Delta z + nUV_x\,\Delta x + nVW_y\,\Delta y - nW_t\,\Delta t\,\Bigr]_z\,\Delta z \\
	&+\ 2\Par{nUV}_{xy}\Delta x\,\Delta y + 2\Par{nVW}_{yz}\Delta y\,\Delta z +  2\Par{nWU}_{zx}\Delta z\,\Delta x\ .
\end{split}
\end{equation}
The substitution of \eqref{eq:n_tt-3D} into \eqref{eq:expansion_3D} leads to the modified equation (to $2^{nd}$ order) for the 3D CCSL-CS:
\begin{equation} \label{eq:3DModifiedEqn}
\begin{split}
	n_t + \Par{un}_x + \Par{vn}_x + \Par{wn}_z\ &=\
	\frac{\Delta x}{2\Delta t}\,\left[\Par{1-U}\Par{nU}_{x}\Delta x\,+\,nVU_y\,\Delta y\,+\,nWU_z\,\Delta z\,+\,nU_t\,\Delta t\,\right]_x\\
	&\ +\frac{\Delta y}{2\Delta t}\,\left[\Par{1-V}\Par{nV}_{y}\Delta y\,+\,nWV_z\,\Delta z\,+\,nUV_x\,\Delta x\,+\,nV_t\,\Delta t\,\right]_y\\
	&\ +\frac{\Delta z}{2\Delta t}\,\left[\Par{1-W}\Par{nW}_{z}\Delta z\,+\,nUW_x\,\Delta x\,+\,nVW_y\,\Delta y\,+\,nW_t\,\Delta t\,\right]_z\\
	&\ +\ H.O.T.
\end{split}
\end{equation}
At this point we can assume that $u$, $v$ and $w$ are obtained by a first order perturbation to the physical velocity:
\begin{equation*}
	\left\{\begin{matrix} u = u_0 + u_1 \\ v = v_0 + v_1 \\ w = w_0 + w_1 \end{matrix}\right.
	\hspace{1cm}\text{or}\hspace{1cm}
	\left\{\begin{matrix} U = U_0 + U_1 \\ V = V_0 + V_1 \\ W = W_0 + W_1 \end{matrix}\right.
\end{equation*}
so that, after substitution into \eqref{eq:3DModifiedEqn}, the corrections are obtained by imposing the compensation of the first order errors. For the general case with no restriction on the sign of the velocity components, one gets:
\begin{subequations} \label{eq:3D_corrections}
\begin{align}
	U_1 &= \left[\Par{|U_0|-U_0^2}\dfrac{\Par{nU_0}_{x}}{nU_0}\right]\dfrac{\Delta x}{2}\,+\,\Bigl[V_0U_{0y}\Bigr]\,\dfrac{\Delta y}{2}\,
	+\,\Bigl[W_0U_{0z}\Bigr]\,\dfrac{\Delta z}{2}\,+\,\Bigl[U_{0t}\Bigr]\,\dfrac{\Delta t}{2} \\
	V_1 &= \left[\Par{|V_0|-V_0^2}\dfrac{\Par{nV_0}_{y}}{nV_0}\right]\dfrac{\Delta y}{2}\,+\,\Bigl[W_0V_{0z}\Bigr]\,\dfrac{\Delta z}{2}\,
	+\,\Bigl[U_0V_{0x}\Bigr]\,\dfrac{\Delta x}{2}\,+\,\Bigl[V_{0t}\Bigr]\,\dfrac{\Delta t}{2} \\
	W_1 &= \left[\Par{|W_0|-W_0^2}\dfrac{\Par{nW_0}_{z}}{nW_0}\right]\dfrac{\Delta z}{2}\,+\,\Bigl[U_0W_{0x}\Bigr]\,\dfrac{\Delta x}{2}\,
	+\,\Bigl[V_0W_{0y}\Bigr]\,\dfrac{\Delta y}{2}\,+\,\Bigl[W_{0t}\Bigr]\,\dfrac{\Delta t}{2}
\end{align}
\end{subequations}
and the correction to be applied to the final position of the MC is simply 
\begin{equation*}
	\delta\mathbf{r}\ =\ \begin{pmatrix} \delta x \\ \delta y \\ \delta z \end{pmatrix}\
	=\ \begin{pmatrix} U_1\Delta x \\ V_1\Delta y \\ W_1\Delta z \end{pmatrix}\ .
\end{equation*}
When the velocity field and the density are smooth functions of space and time, the corrections above lead to a simple and efficient $2^{nd}$ order version of the 3D CCSL-CS, for a generic velocity field. As usual, the scheme is conservative and positivity preserving.

%% file: MonotonicVariations.tex
\section{Monotonic variations} \label{sec:MonotonicOption}

For a constant velocity field $u_0$, the first order version of the 1D CCSL-CS guarantees the solution $n\Par{x,t}$ will conserve the monotonicity of the initial conditions. Unfortunately, this is not true for the third order version. In fact, wherever the advected function $n(x)$ happens not to be differentiable in space, the finite difference approximations in \eqref{eq:U1} and \eqref{eq:U2} become meaningless and the estimates for $U_1$ and $U_2$ are incorrect. At best, the accuracy of the scheme drops to first order; more likely, $U_1$ and $U_2$ are overestimated and $O(1)$. As a consequence, Gibbs' oscillations would appear, with an amplitude that does not decrease on refining the mesh. In such a situation the scheme would probably retain a first-order accuracy in the $L_2$ norm, but \emph{it cannot converge to the correct solution} in the $L_{\infty}$ norm.

\subsection*{Limiters}

In order to make the third order CCSL-CS monotonic, i.e.\ non-oscillating, it should be noticed that the spurious oscillations are due to the corrections $U_1$ and $U_2$. Hence, a natural approach is to reduce the magnitude of these corrections with a \emph{limiter function} wherever the conditions are favorable for spurious oscillations to arise. Basically, in order to preserve monotonicity, the limiter function is supposed to reduce the scheme to first order in the proximity of a discontinuity in $n_x$. In practice, it happens that $U_1$ and $U_2$ are reduced also in regions where no such limiting is needed: a typical long-time effect is the artificial flattening of smooth local maxima and minima. Several limiter functions were obtained by adapting the flux limiters commonly used in high order Total Variation Diminishing (TVD) schemes \cite{Boris1973,VanLeer1974_2,Harten1983,Sweby1984}. Here, the general implementation is based on the slight modification of \eqref{eq:U_expansion} into
\begin{equation*}
	U = U_0 + \phi\Par{r}\left[U_1 + U_2\right]
\end{equation*}
where $\phi\Par{r}$ is the limiter function and $r$ is the ratio of successive gradients on the solution mesh, with a slightly modified definition in order to assure that $-1\leq r\leq 1$:
\begin{equation*}
	r = \min\left[\frac{n_i-n_{i-1}}{n_{i+1}-n_i},\frac{n_{i+1}-n_i}{n_i-n_{i-1}}\right].
\end{equation*}
In the tests performed so far, the best results were obtained using the Superbee limiter designed by Roe \cite{Roe1986} in 1986
\begin{equation*}
	\phi\Par{r} = \max\left[0,\min\Par{2r,1}\right],
\end{equation*}
where the original simple formula is further simplified here because $r\leq 1$.

\theoremstyle{definition}
\newtheorem*{defn}{Definition}  
\subsection*{Non-polynomial reconstruction}

Taking a slightly different point of view, the finite difference approximations in \eqref{eq:U1} and \eqref{eq:U2} are evaluated from a local quadratic polynomial interpolation of the discrete function $\{n_i\}$ between three equally spaced grid points: once the interpolating polynomial is obtained, it is analytically differentiated in order to give an approximation for the first and second derivatives of $n\Par{x}$ at the center point.
In general, it is well-known that polynomial interpolations are not suited to reconstruct `non-smooth' functions on a uniform grid, and hence it is no surprise that spurious oscillations are triggered whenever there is a discontinuity in the solution.

Following naturally from these considerations is the idea of using non-polynomial functions to reconstruct the density function inside a cell.
In \cite{Artebrant2006} a third order non-oscillating reconstruction technique is outlined, which uses only three grid points (resulting in an ideally compact stencil), and that does not need limiters.
Such a method is based on a \emph{Local Double Logarithmic Reconstruction} (LDLR) of the form
\begin{equation*}
	r_0\Par{x} \simeq A + B\log\Par{x+C} + D\log\Par{x+E},
\end{equation*}
within cell $C_{0}$.
As described is more detail later on, the five parameters in last equation are obtained by imposing conservation, formal third order accuracy, symmetry, and Local Variation Boundedness (LVB).
The complete form of the reconstructing function is
\begin{equation} \label{eq:LDLR}
	r_0\Par{x} = \overline{n}_{0} + \phi_{0}\Par{x} - \frac{1}{\dx}\int_{C_{0}}\phi_{0}\Par{\xi}\,d\xi
\end{equation}
where $\overline{n}_{0}$ is the average value of $n\Par{x}$ within cell $C_{0}$, and $\frac{1}{\dx}\int_{C_{0}}\phi_{0}\Par{\xi}\,d\xi$ represents the mean value of the function $\phi_{0}\Par{x}$, which is defined as:
\begin{equation} \label{eq:LDLR_phi}
	\phi_{0}\Par{x} = -\frac{c\,\dx}{a}\log \left[ \Par{x-x_0}-\frac{\dx}{2}\Par{\frac{2}{a}-1} \right] - \frac{d\,\dx}{b}\log \left[ \Par{x-x_0}-\frac{\dx}{2}\Par{\frac{2}{b}-1} \right] .
\end{equation}
For any choice of the coefficients $a$, $b$, $c$ and $d$, the reconstruction \eqref{eq:LDLR} is conservative, since
\begin{equation}
	\frac{1}{\dx}\int_{C_0}r_0\Par{x}\,dx \equiv \overline{n}_{0}\ .
\end{equation}
The requirement of third order accuracy requires the reconstruction to match the second order finite difference approximations to the derivative $u'\Par{x}$ at the cell faces:
\begin{align*}
	r'_{0}\Par{x_{0}-\frac{\dx}{2}} &= \frac{\overline{u}_{0} - \overline{u}_{-1}}{\dx} = \delta_{1}\ , \\
	r'_{0}\Par{x_{0}+\frac{\dx}{2}} &= \frac{\overline{u}_{1} - \overline{u}_{0}}{\dx} = \delta_{2}\ ,
\end{align*}
which permits one to solve for the parameters $c$ and $d$ as functions of $a$, $b$, $\delta_{1}$ and $\delta_{2}$:
\begin{align}
	\label{eq:LDLR_c}
	c &= \frac{\Par{a-1}\left[\delta_{2}\Par{1-b}-\delta_{1}\right]}{b-a}\ , \\
	\label{eq:LDLR_d}
	d &= \delta_{1}-c\ .
\end{align}
The reconstruction should be symmetric, in the sense that if $r'_{0}\Par{x_{0}-\dx/2} = -r'_{0}\Par{x_{0}+\dx/2}$, then $r'_{0}\Par{x_{0}}$ should be zero.
This permits one to find $b$ as a function of $a$,
\begin{equation} \label{eq:LDLR_b}
	b = \frac{a}{a-1}\ ,
\end{equation}
and hence all the 3 parameters $b$, $c$ and $d$ that show up in \eqref{eq:LDLR_phi} are univocally determined once that $a$ is known.
To construct a reconstruction procedure of sufficiently low variation Marquina's concept of \emph{Local Variation Boundedness} is used:
\begin{defn}[Local Variation Bounded \cite{Marquina1994}]
	The local variation of a function $f\Par{x}$ in a cell $C_{i}$ is given by $LV\Par{f_{i}}=TV\Par{f}|_{C_{i}}$.
	The function is Local Variation Bounded (LVB) in $C_{i}$ if $LV\Par{f_{i}} = O(\dx)$, where $\dx$ is the cell size.
\end{defn}
An expression for $a\Par{\delta_{1},\delta_{2}}$ which ensures that the reconstruction \eqref{eq:LDLR_phi} be a well defined smooth LVB function is \cite{Artebrant2006}
\begin{equation} \label{eq:LDLR_a}
	a\Par{\delta_1,\delta_2} = \Par{1-\textmd{TOL}}\Par{1+\textmd{TOL}-\frac{2\, |\delta_{1}|^{q}\, |\delta_{2}|^{q} + \textmd{TOL}}{|\delta_{1}|^{2q} + |\delta_{2}|^{2q} + \textmd{TOL}}}\ ,
\end{equation}
where $\textmd{TOL} = 0.1\dx^{q}$ and $q=1.4$.
Once that $a$ is determined from \eqref{eq:LDLR_a}, $b$, $c$ and $d$ are obtained from \eqref{eq:LDLR_b}, \eqref{eq:LDLR_c} and \eqref{eq:LDLR_d}, respectively.
For the purposes of the third-order CCSL-CS, the first and second derivatives of the reconstruction \eqref{eq:LDLR} are needed, which only need to be evaluated for $x=x_{0}$, giving the following compact formulas:
\begin{subequations}
\begin{align*} \label{eq:LDLR_derivs} 
	r'_{0}\Par{x_0} &= \dfrac{2\,c}{2-a} + \dfrac{2\,d}{2-b}\ ,  \\
	r''_{0}\Par{x_0} &= -\dfrac{1}{\dx}\left[\dfrac{4\,a\,c}{\Par{2-a}^2} + \dfrac{4\,b\,d}{\Par{2-b}^2}\right]\ .
\end{align*}
\end{subequations}
The most demanding part of the algorithm is the evaluation of the coefficient $a$ by means of \eqref{eq:LDLR_a}, since it requires the exponentiation of $|\delta_1|$ and $|\delta_2|$ by the real power $q$, operation which is usually much slower than taking the natural logarithm or exponential.
The number of exponentiations can be reduced to one if the \emph{smoothness parameter} $\theta$ is used,
\begin{equation*} \label{eq:LDLR_theta} 
	\theta_0 = \frac{\dx\,\min\Par{|\delta_1|,|\delta_2|}+\textmd{TOL}}{\dx\,\max\Par{|\delta_1|,|\delta_2|}+\textmd{TOL}}
\end{equation*}
where \textmd{TOL} is in this case a small number that depends on the machine precision $\epsilon$; robustness and accuracy can be achieved by choosing $\textmd{TOL} = \epsilon^{3/4}$.
The expression for $a$ can be reformulated as a function of $\theta_0$ alone,
\begin{equation*} \label{eq:LDLR_a(theta)} 
	a\Par{\theta_0} = \Par{1-\textmd{TOL}}\Par{1+\textmd{TOL}-\frac{2\Par{\theta_0}^{q}}{1+\Par{\theta_0}^{2q}}}\ ,
\end{equation*}
so that only one exponentiation to a real power is required.
In the numerical tests performed, a considerable speedup was obtained by creating a look-up table for $a\Par{\theta_0}$ in preprocessing.

%% file: NumericalTests.tex
\section{Numerical tests} \label{sec:NumericalTests}

The order-of-convergence estimates that were carried out in the preceding sections were based exclusively on what we called the `modified equation error' (MEE), which is the difference between the model equation and the modified equation of the scheme.
Accordingly, we labeled a scheme as `first order' or `third order' depending on the form of the MEE.
This can be misleading, because we are interested in approximating to high order not just the model equation, but the solution itself.
In fact, the Modified Equation that is solved by the numerical scheme is a \emph{singular perturbation} to the model equation, and as such it may lead to `spurious' features that were not present before.
Whether or not such features are triggered depends mainly on the initial density profile and on the boundary conditions.

Following from the above considerations, the primary purpose of this section is to assess the Global Error (GE) of the scheme, which is the actual difference between the \emph{exact solution} and the \emph{approximated solution}.
The GE is quantified a-posteriori in the $L_{\infty}$, $L_2$ or $L_1$ norm, depending on what is more appropriate, and convergence analyses are performed by successively decreasing the mesh size. A limited number of test-cases are considered, both in 1D and 2D.

In order to show the capabilities of the scheme in a `real-world' application, the final part of this section is devoted to a 3D-3V kinetic simulation: the expansion of a rarefied neutral gas injected through a cylinder into vacuum will be shown, for different Knudsen regimes.

\subsection*{1D advection equation}

As the simplest test case, the third order CCSL-CS is applied to the solution of the 1D continuity equation \eqref{eq:Continuity1D} in the hypothesis of constant and uniform velocity.
The solution $n\Par{x,t}>0$ is sought over the domain $\Par{x,t}\in\left[0,L\right]\times\left[0,T\right]$ with periodic boundary conditions in space $n\Par{0,t} \equiv n\Par{L,t}$, and initial conditions $n\Par{x,0} = n_0\Par{x}$.
The independent variables are normalized by introducing $\xi = x/L$ and $\tau = t/T_p$, where $T_p = L/u$ is the characteristic flow time.
Under these assumptions, we will numerically solve
\begin{equation} \label{eq:test1D_norm}
	n_{\tau} + n_{\xi} = 0
\end{equation}
on the domain $\Par{\xi,\tau}\in\left[0,1\right]\times\left[0,1\right]$ (i.e., $T=T_p$ is chosen). The analytic solution to \eqref{eq:test1D_norm} is $n\Par{\xi,\tau} = n_0\Par{\xi-\tau}$; due to the periodic boundary conditions, the final solution coincides with the initial conditions: $n\Par{\xi,1} = n_0\Par{\xi}$.
The application of the third order CCSL-CS to \eqref{eq:test1D_norm} comes from the usual CS remapping rule, which is used after moving the cell a distance $U\Delta\xi$, where $U = U_0 + U_1 + U_2$.
We have $U_0 = u\,\Delta t/\Delta x = \Delta\tau/\Delta\xi$, and it will be imposed here that $|U_0| < 1$.
From \eqref{eq:U1} and \eqref{eq:U2}, the first and second order corrections in the case of uniform and constant (but not necessarily positive) $U_0$ are:
\begin{align*}
	U_1 &= \frac{\Delta\xi}{2}\Par{|U_0|-U_0^2}\frac{n_{\xi}}{n} \\
	U_2 &= \frac{\Delta\xi^2}{12}\Par{U_0-3|U_0|U_0+2U_0^3}\frac{n_{\xi\xi}}{n}
\end{align*}
which will be approximated using the finite difference expressions
\begin{subequations}
\begin{align}
	\label{eq:FD_1st}
	\left.\frac{1}{n}\frac{\partial n}{\partial\xi}\right|_{\xi=\xi_i} &=
	\left.\frac{\partial\Par{\log{n}}}{\partial\xi}\right|_{\xi=\xi_i} =
	\frac{1}{\Delta\xi}\log\Par{\frac{n_{i+1}}{n_{i-1}}} + O\Par{\Delta\xi^2} \\
	\label{eq:FD_2nd}
	\left.\frac{1}{n}\frac{\partial^2 n}{\partial\xi^2}\right|_{\xi=\xi_i} &=
	\frac{4}{\Delta\xi^2}\frac{n_{i+1}-2n_i+n_{i-1}}{n_{i+1}+2n_i+n_{i-1}}
	 + O\Par{\Delta\xi^2}
\end{align}
\end{subequations}
but different choices can be made, as long as a second order approximation is used for \eqref{eq:FD_1st} and a first order approximation is used for \eqref{eq:FD_2nd}.

\begin{figure}[!ht]
	\includegraphics[width=\textwidth]{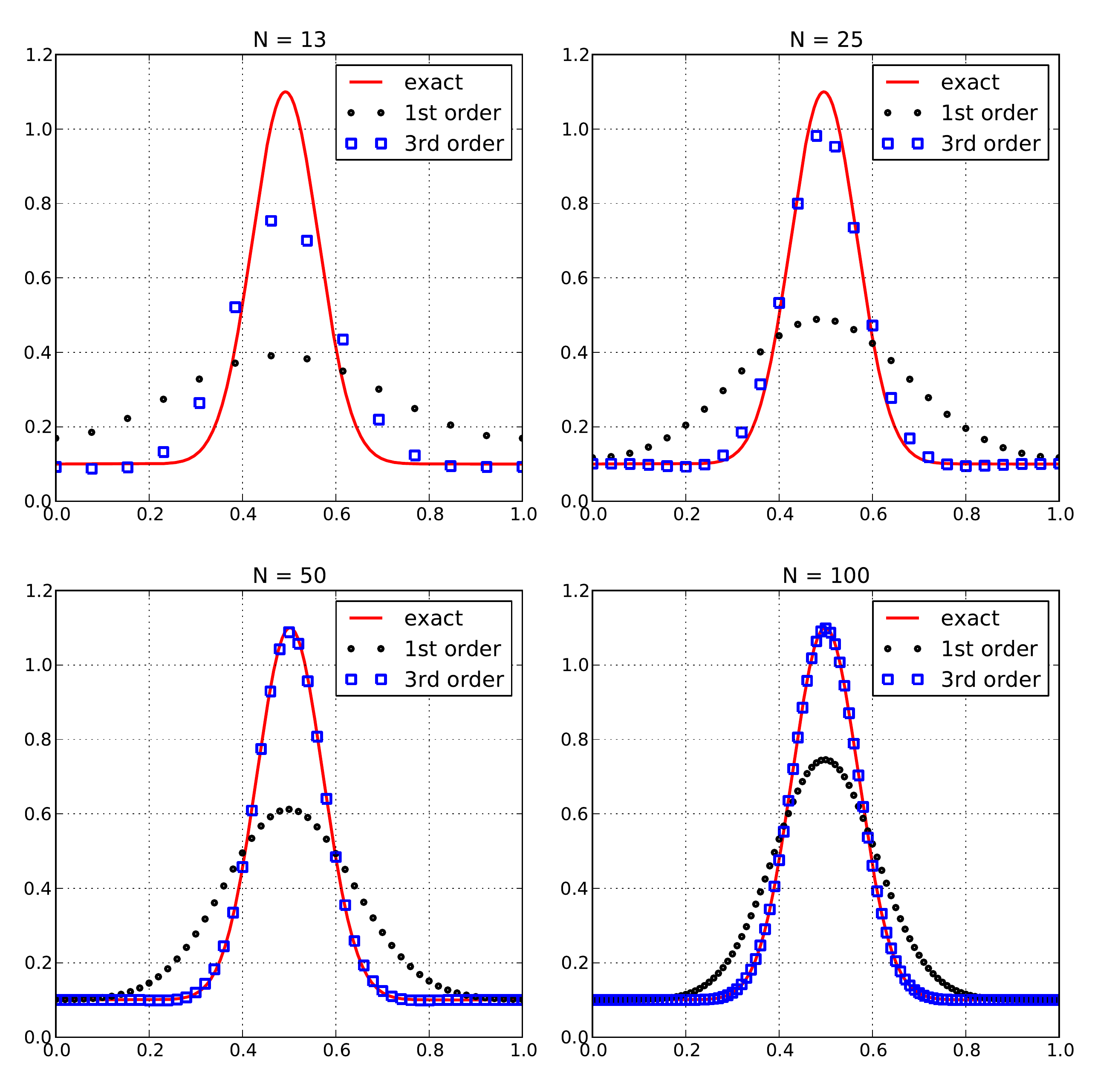}
	\caption{1D advection with Gaussian initial profile: final solution at time $\tau=1.0$ for increasing number N of spatial subdivisions.}
	\label{fig:FinalSolution1D}
\end{figure}

\begin{figure}[!ht]
	\centering
	\includegraphics[width=0.75\textwidth]{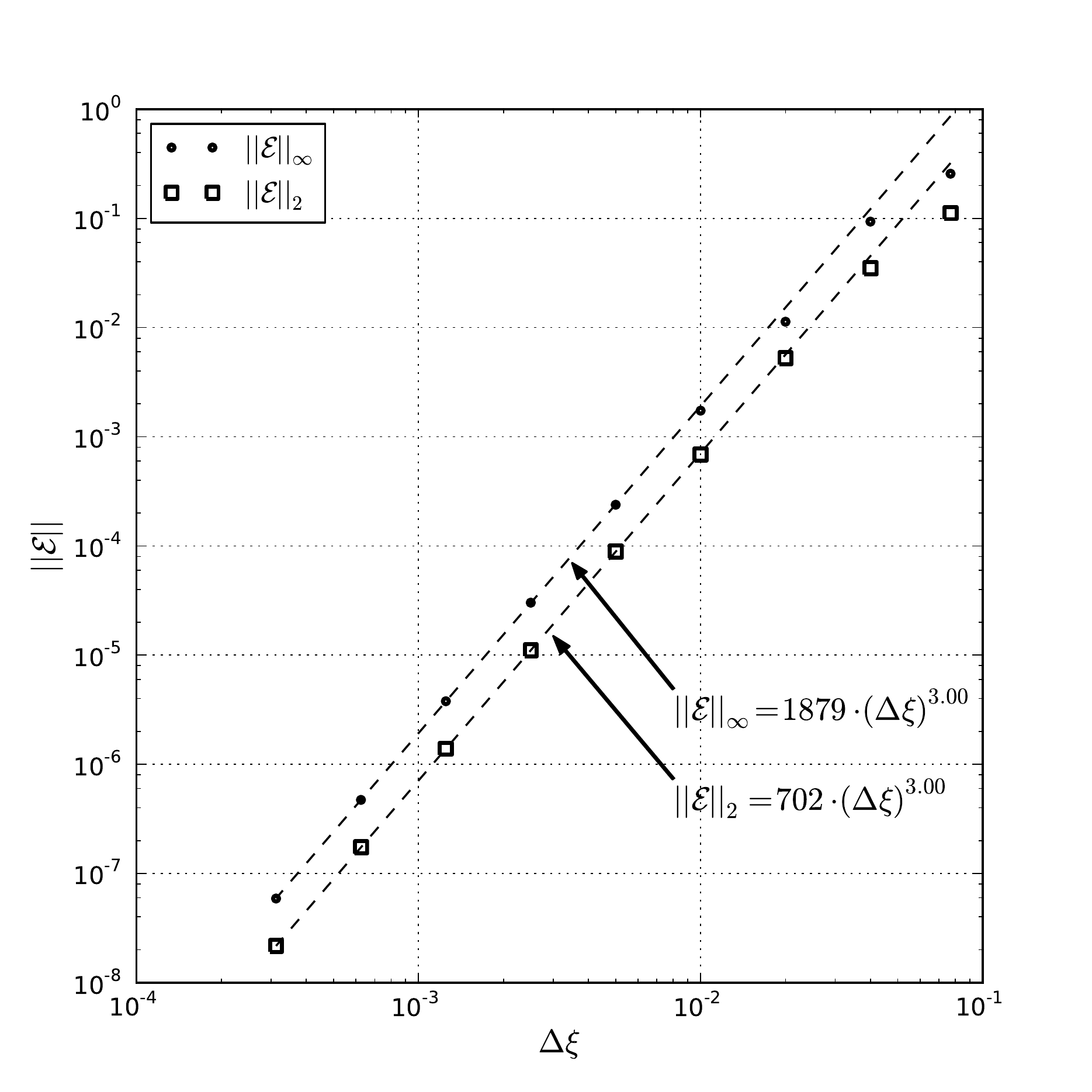}
	\caption{Convergence analysis for 1D advection with Gaussian initial profile. The L$_\infty$ and L$_2$ norms of the error at the end of the simulation ($\tau=1.0$) are shown for decreasing mesh-size $\Delta\xi$.}
	\label{fig:Convergence1D}
\end{figure}

As a first test case, a Gaussian profile centered at $\xi=0.5$ is given as initial condition,
\begin{equation*}
	n\Par{\xi,0} = 0.1 + \exp{\left[-\Par{\frac{\xi-0.5}{0.1}}^2\right]}
\end{equation*}
and then evolved in time for $\tau\in\left[0,1\right]$. A Courant parameter $U_0 = 0.3$ is chosen. The final solution obtained with the third order CCSL-CS is compared qualitatively with the original first order version in Fig.\ \ref{fig:FinalSolution1D}, for different numbers $N$ of subdivisions in space ($\Delta\xi = 1/N$). The $L_2$ and $L_{\infty}$ norms of the error for the third order scheme are also computed by doubling the number of subdivisions up to 3200 cells. The results are summarized by Fig.\ \ref{fig:Convergence1D}, which confirms that the error of the scheme decreases as $\Par{\Delta\xi}^3$ in both the norms; it is most noticeable that the asymptotic convergence behavior of the scheme is reached very soon, starting from $N\approx 30$.

\begin{figure}[!ht]
	\includegraphics[width=\textwidth]{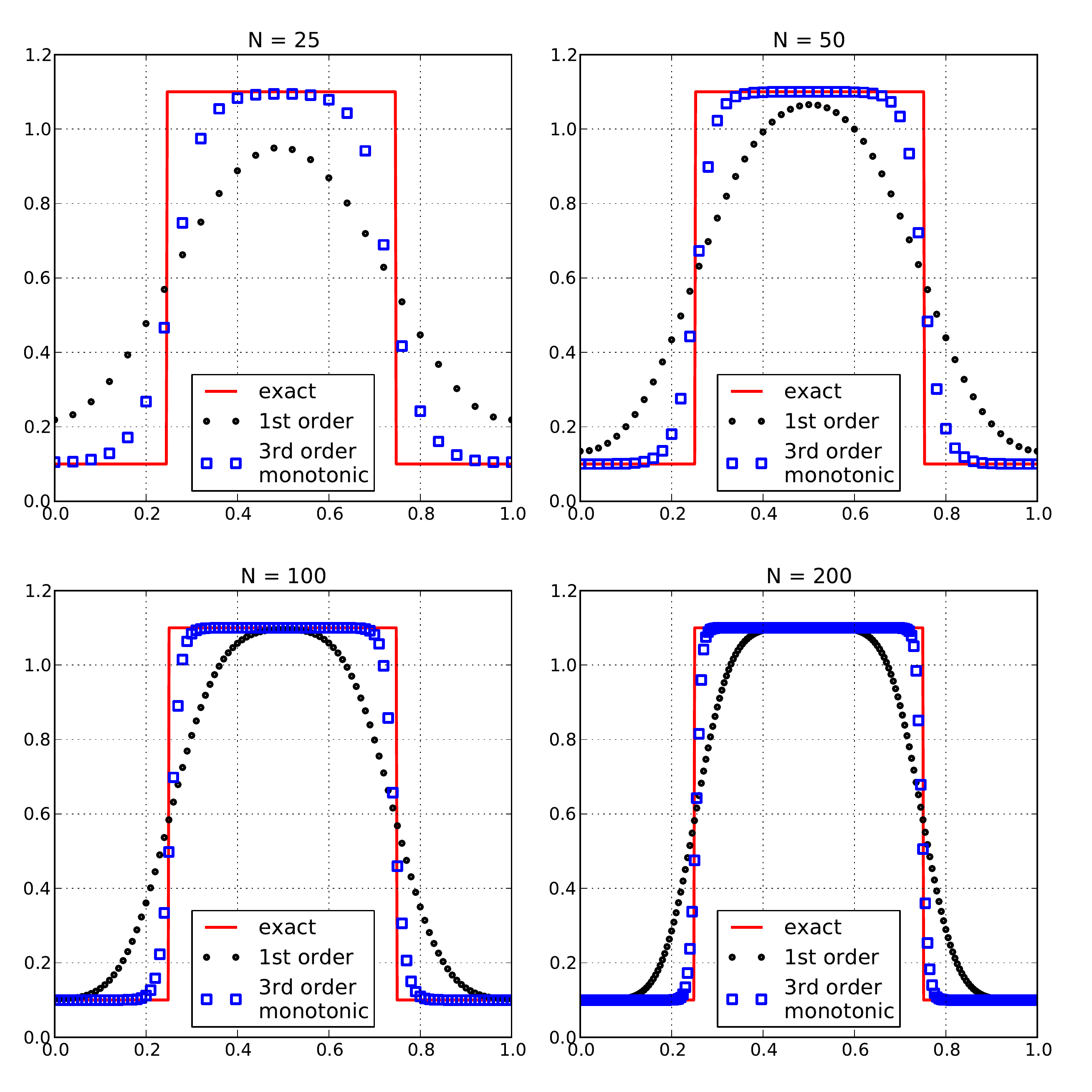}
	\caption{1D advection with rectangular initial profile: final solution at time $\tau=1.0$ for increasing number N of spatial subdivisions.}
	\label{fig:FinalSolution1D_rect}
\end{figure}

\begin{figure}[!ht]
	\centering
	\includegraphics[width=0.75\textwidth]{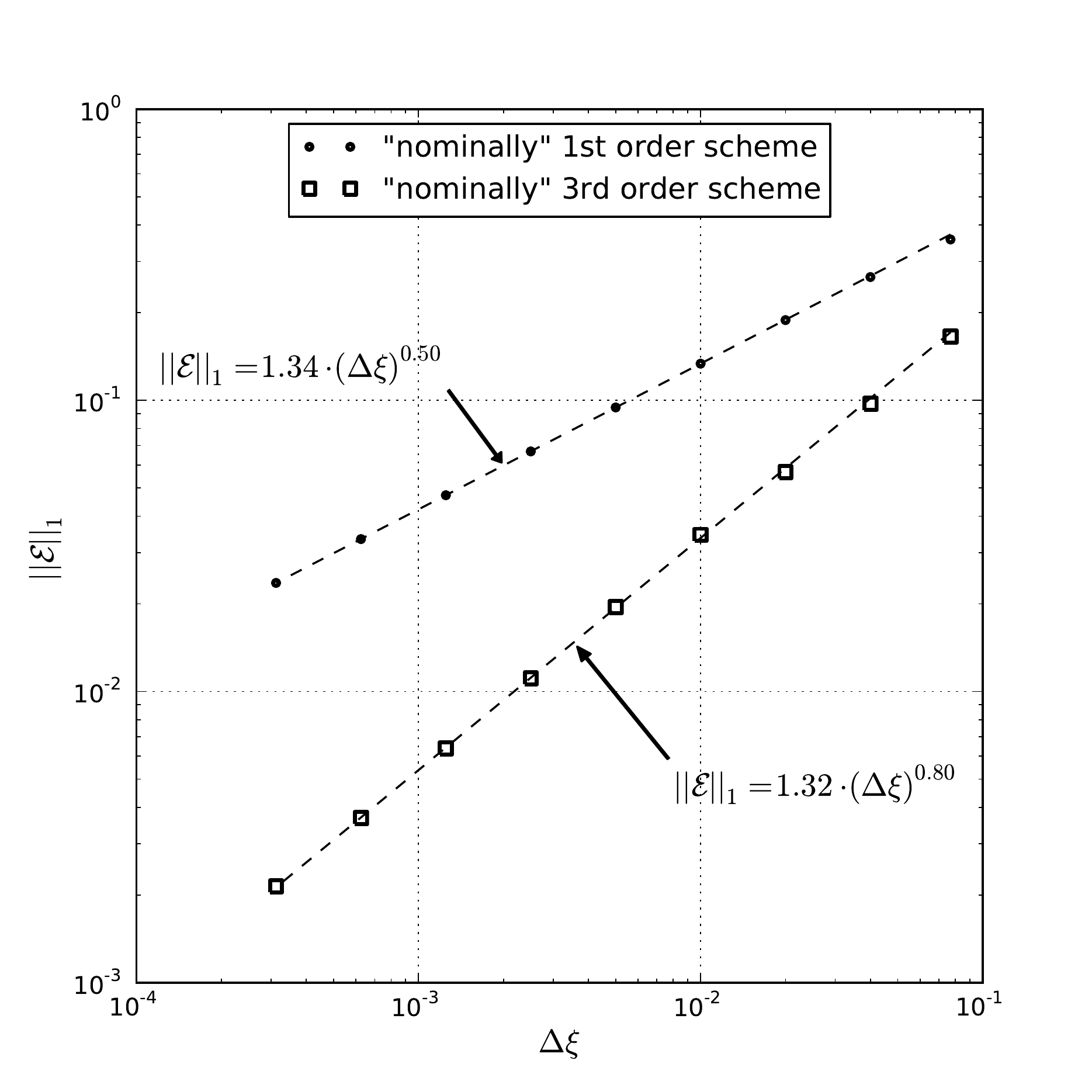}
	\caption{Convergence analysis for 1D advection with rectangular initial profile. The L$_1$ norm of the error at the end of the simulation ($\tau=1.0$) is shown for decreasing mesh-size $\Delta\xi$.}
	\label{fig:Convergence1D_rect}
\end{figure}

As a second test case, a rectangular profile is given as initial condition:
\begin{equation*}
	n\Par{\xi,0} =
	\begin{cases}
	1.1 & \text{if $0.25 < \xi < 0.75$}, \\
	0.1 & \text{otherwise}.
	\end{cases}
\end{equation*}
As before, $\tau\in\left[0,1\right]$ and $U_0 = 0.3$, and a series of identical simulations is performed using an increasingly fine mesh. Since the initial density profile is not `smooth', the $3^{rd}$ order CCSL-CS undergoes non-physical oscillations in the regions close to the discontinuities. This effect does not compromise the positive-definiteness of the scheme, which is inherited from the CS remapping rule, but it considerably deteriorates the solution. Hence, the non-oscillatory option described in Section \ref{sec:MonotonicOption} is implemented here, and Fig.\ \ref{fig:FinalSolution1D_rect} gives a qualitative picture of the improvement upon the standard $1^{\text{st}}$ order scheme.
As is common practice in the case of discontinuous solutions, a convergence analysis based on the $L_1$ norm of the error is presented in Fig.\ \ref{fig:Convergence1D_rect}: as expected, both schemes exhibit sub-linear convergence \cite{Banks2008}.

The GE of the (nominally) $1^{\text{st}}$ order scheme goes as $\Par{\Delta\xi}^{0.5}$, a result obtained a-priori in \cite{LeVeque1992} (pg.\ 121). The GE of the (nominally) $3^{rd}$ order scheme is proportional to $\Par{\Delta\xi}^{0.8}$, which is slightly better than the expected $\Par{\Delta\xi}^{0.75}$ for the non-limited case: in fact, in \cite{Banks2008} the order of convergence for unlimited stable schemes is established as $\Par{\Delta\xi}^{p/\Par{p+1}}$ where $p$ is the order of the approximation for smooth flows.

\begin{figure}[!ht]
\includegraphics[width=\textwidth]{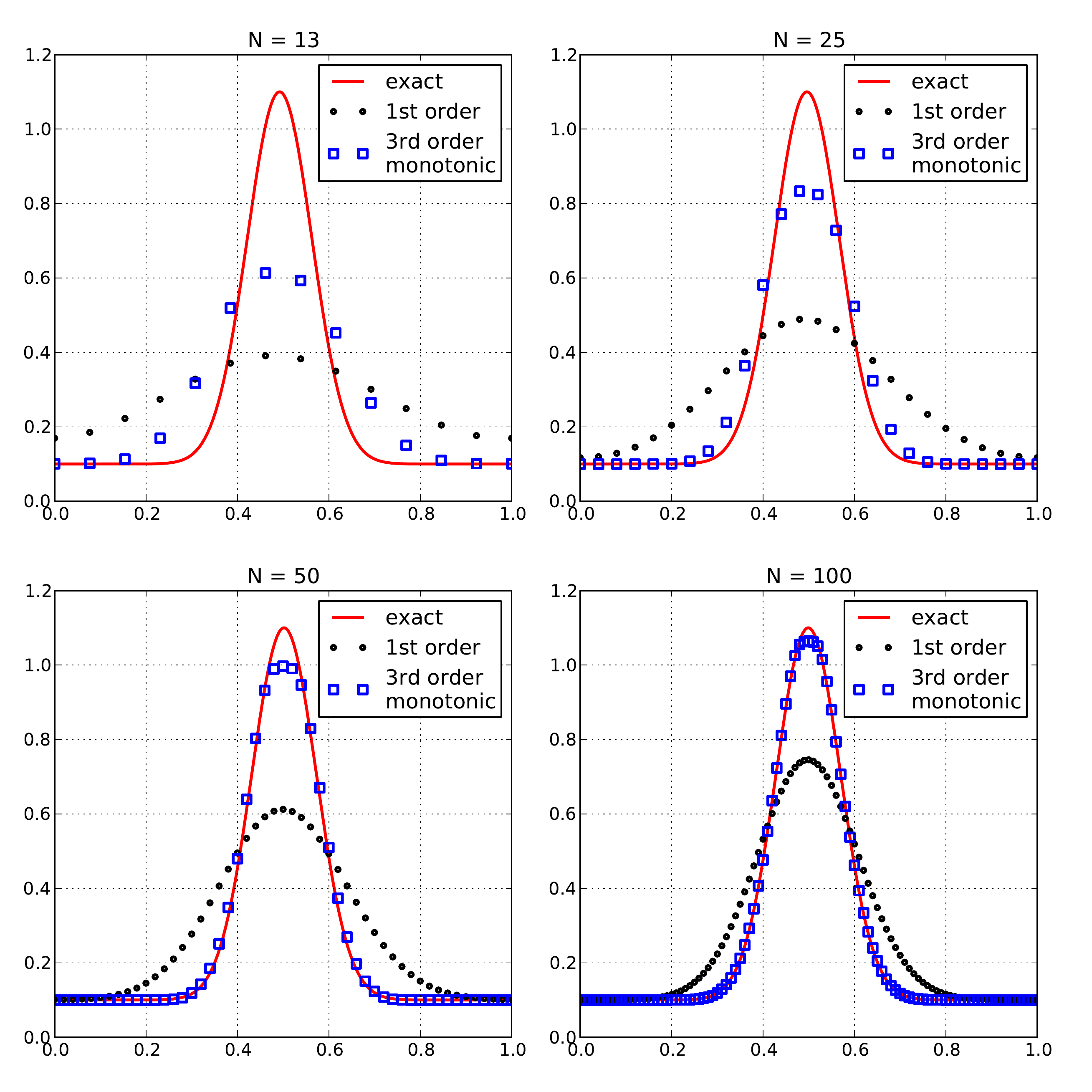}
\caption{1D advection with Gaussian initial profile: final solution at time $\tau=1.0$ for increasing number N of spatial subdivisions.
A non-oscillating version of the high-order CCSL-CS is employed, which uses a Local Double Logarithmic Reconstruction (LDLR) for evaluating the spatial derivatives of the density.}
\label{fig:FinalSolution1D_Gaussian_LDLR}
\end{figure}

\begin{figure}[!ht]
\centering
\includegraphics[width=0.75\textwidth]{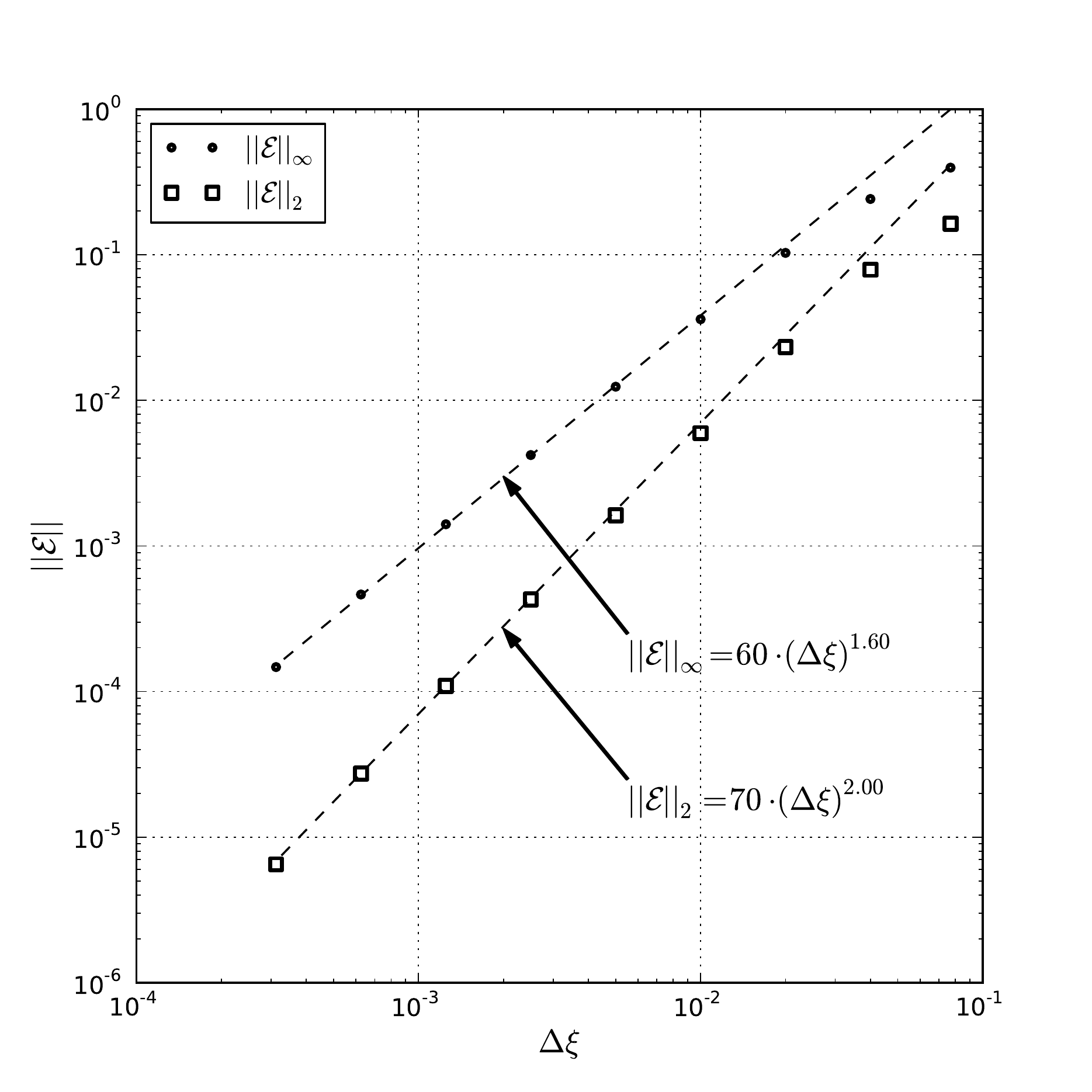}
\caption{
Convergence analysis for 1D advection with Gaussian initial profile, for the non-oscillating version of the high-order CCSL-CS, which uses a Local Double Logarithmic Reconstruction (LDLR) of the derivatives.
The L$_\infty$ and L$_2$ norms of the error at the end of the simulation ($\tau=1.0$) are shown for decreasing mesh-size $\Delta\xi$.}
\label{fig:Convergence1D_Gaussian_LDLR}
\end{figure}

One may want to investigate the behavior of the monotonic version of the high-order CCSL-CS when `smooth' initial conditions are given, in order to assess how the performance degrades with respect to the non-monotonic version. 
For this purpose, the first test case (with a Gaussian profile as initial conditions) is repeated using the Local Double Logarithmic Reconstruction described in Section \ref{sec:MonotonicOption}.
The results are shown in Fig.\ \ref{fig:FinalSolution1D_Gaussian_LDLR}, for an increasing number of subdivisions in space: as typical of monotonicity-preserving schemes, the peak of the Gaussian profile is flattened by the numerical diffusion that is introduced in the non-monotonic regions of the density profile.
Accordingly, the accuracy of the scheme is significantly affected: instead of the cubic convergence shown in Fig.\ \ref{fig:Convergence1D}, the monotonic version of the high-order CCSL-CS achieves quadratic convergence in the $L_2$ norm, and only sub-quadratic convergence in the $L_\infty$ norm (see Fig.\ \ref{fig:Convergence1D_Gaussian_LDLR}).

Given the higher accuracy of the non-monotonic scheme, and considered the fact that it is inherently positivity-preserving, the monotonic version is only preferred in those situations where spurious oscillations in the solution are expected.

\subsection*{2D continuity equation with rotating velocity field}

As a test case with non-uniform velocity, the continuity equation is solved on a square domain with a given rotating velocity field. The equation to be solved is:
\begin{equation*}
	\partial_t\, n + \partial_x\Par{u\,n} + \partial_y\Par{v\,n} = 0
\end{equation*}
on the square domain $\Par{x,y}\in\left[0,L\right]\times\left[0,L\right]$ and in the time interval $t\in\left[0,T\right]$. The two components of the velocity vector are
\begin{align*}
	u &= -\omega \Par{y-L/2} \\
	v &= +\omega \Par{x-L/2}\ ,
\end{align*}
where $\omega = 2\pi/T$ [rad/s] is the angular frequency of the rotating field. After normalization of the spatial and temporal variables with respect to $L$ and $T$, one gets:
\begin{equation}\label{eq:test2D_norm}
	\partial_{\tau}\, n + \partial_{\xi}\Par{\overline{u}\,n} + \partial_{\eta}\Par{\overline{v}\,n} = 0
\end{equation}
where $\Par{\xi,\eta}\in\left[0,1\right]\times\left[0,1\right]$, $\tau\in\left[0,1\right]$, and the normalized velocities were introduced:
\begin{subequations}\label{eq:2D_normalized_velocities}
\begin{align}
	\overline{u} &= u\,T/L = -2\pi \Par{\eta-0.5} \\
	\overline{v} &= v\,T/L = +2\pi \Par{\xi-0.5}\ .
\end{align}
\end{subequations}
The application of the second order CCSL-CS to \eqref{eq:test2D_norm} comes from the usual CS remapping rule, which is used after moving the cell a distance $\Delta\vect{r}=\left[U\Delta\xi,V\Delta\eta\right]$, where $U = U_0 + U_1$ and $V = V_0 + V_1$. With the new variables we have $U_0 = u\,\Delta t/\Delta x = \overline{u}\,\Delta\tau/\Delta\xi$ and $V_0 = v\,\Delta t/\Delta y = \overline{v}\,\Delta\tau/\Delta\eta$. 
The same number of subdivisions is employed along both the x and y axes, so that $\Delta\eta=\Delta\xi$. The corrections $U_1$ and $V_1$ are evaluated using second order finite difference approximations to \eqref{eq:3D_corrections}:
\begin{align*}
	\Par{U_1}_{i,j}\ &=\ \frac{1}{2}\,\Par{|U_0|-U_0^2}_{i,j}
	\frac{\Par{n\,U_0}_{i+1,j}-\Par{n\,U_0}_{i-1,j}}
	     {\Par{n\,U_0}_{i+1,j}+\Par{n\,U_0}_{i-1,j}}\ +\
	\frac{1}{4}\Par{V_0}_{i,j}\Bigl[\Par{U_0}_{i,j+1}-\Par{U_0}_{i,j-1}\Bigr] \\
	\Par{V_1}_{i,j}\ &=\ \frac{1}{2}\,\Par{|V_0|-V_0^2}_{i,j}
	\frac{\Par{n\,V_0}_{i,j+1}-\Par{n\,V_0}_{i,j-1}}
	     {\Par{n\,V_0}_{i,j+1}+\Par{n\,V_0}_{i,j-1}}\ +\
	\frac{1}{4}\Par{U_0}_{i,j}\Bigl[\Par{V_0}_{i+1,j}-\Par{V_0}_{i-1,j}\Bigr]
\end{align*}
The maximum Courant parameter in the domain is reached at the boundaries and it is fixed at 1, i.e.\ $\left|U_0\right|_{\text{max}} = \left|\,\overline{u}_{\text{max}}\Delta\tau/\Delta\xi\right| = 1$, and $\left|V_0\right|_{\text{max}} = \left|\,\overline{v}_{\text{max}}\Delta\tau/\Delta\eta\right| = 1$.
According to \eqref{eq:2D_normalized_velocities}, the time step is also fixed:
\begin{equation*}
	\Delta\tau = \frac{\Delta\xi}{\pi}\ .
\end{equation*}
The system is evolved for a time $\tau\in\left[0,0.25\right]$, which means that the bell performs one quarter ($\pi/2$) of a full rotation ($2\pi$), during the whole simulation.
A cosine bell profile is given as initial condition:
\begin{equation*}
	n\Par{\xi,\eta,0} =
	\begin{cases}
		0.1 + \dfrac{1}{2}\left[1.0 +\cos\Par{\dfrac{\pi r}{R}}\right] \quad &\text{if $r<R$}\\
		0.1 \quad &\text{otherwise}
	\end{cases}
\end{equation*}
where $r = \sqrt{\Par{\xi-\xi_c}^2+\Par{\eta-\eta_c}^2}$ is the distance from the center $\Par{\xi_c,\eta_c}$ of the bell, and $R$ is the `radius' of the bell.
In the following simulations, the center is chosen at the point $\Par{\xi_c,\eta_c}=\Par{0.5, 0.75}$, and a radius $R=0.2$ is used.

\begin{figure}[!ht]
\centering
\includegraphics[width=\textwidth]{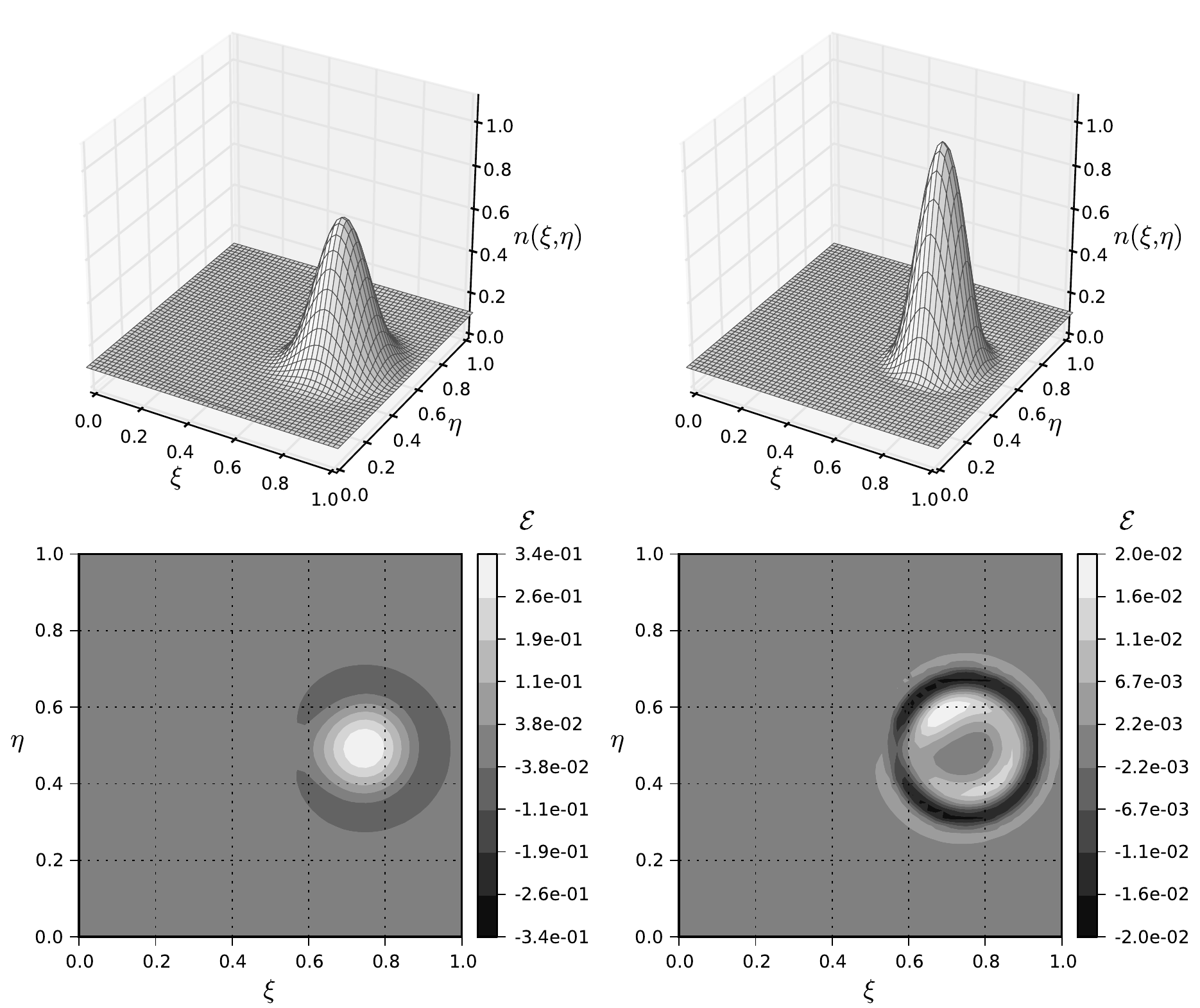}
\caption{
Results for 2D advection with rotating velocity field, for a number of subdivisions $N_x = N_y = 50$.
A cosine bell is given as the initial density profile.
On the top row: final solution evaluated by the $1^{\text{st}}$ order scheme (left) and by the $2^{\text{nd}}$ order scheme (right).
On the bottom row: error for the $1^{\text{st}}$ (left) and the $2^{\text{nd}}$ order scheme (right).}
\label{fig:Results2D_N50}
\end{figure}

\begin{figure}[!ht]
\centering
\includegraphics[width=\textwidth]{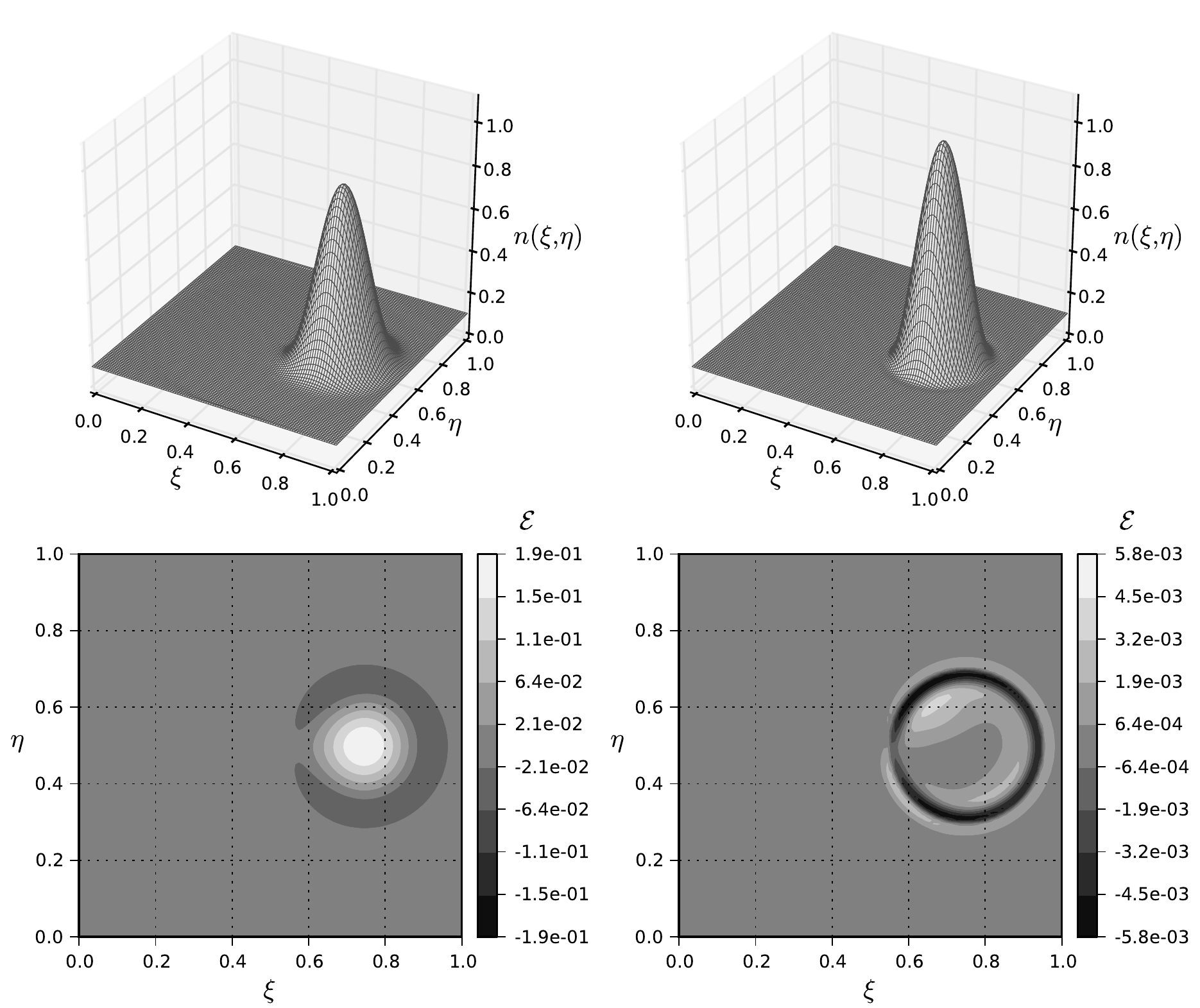}
\caption{
Results for 2D advection with rotating velocity field, for a number of subdivisions $N_x = N_y = 100$.
A cosine bell is given as the initial density profile.
On the top row: final solution evaluated by the $1^{\text{st}}$ order scheme (left) and by the $2^{\text{nd}}$ order scheme (right).
On the bottom row: error for the $1^{\text{st}}$ (left) and the $2^{\text{nd}}$ order scheme (right).}
\label{fig:Results2D_N100}
\end{figure}

\begin{figure}[!ht]
	\centering
	\includegraphics[width=0.75\textwidth]{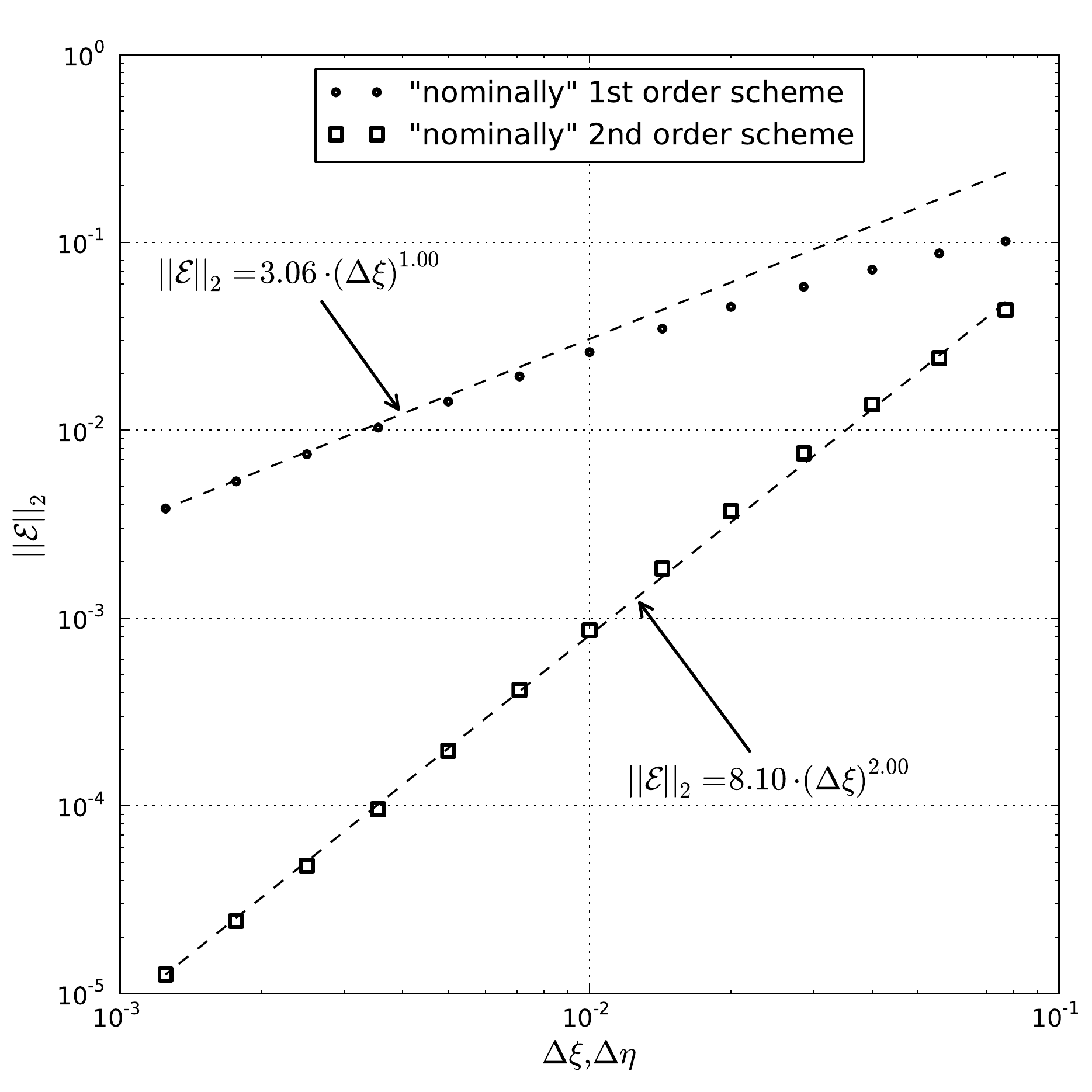}
	\caption{Convergence analysis for 2D advection with rotating velocity field.
	A cosine bell is given as the initial density profile.
	The L$_2$ norm of the error at the end of the simulation ($\tau=0.25$) is shown for decreasing mesh-size $\Delta\xi \equiv \Delta\eta$.}
	\label{fig:Comparison2D}
\end{figure}

The final solution and the final error are shown in Figs.\ \ref{fig:Results2D_N50} and \ref{fig:Results2D_N100}, using respectively 50 and 100 subdivisions along each axis; the use of the $2^{\text{nd}}$ order scheme considerably improves the solution over the original $1^{\text{st}}$ order method.
To complete the analysis, Fig.\ \ref{fig:Comparison2D} shows the asymptotic behavior of the error of the two schemes for decreasing mesh-size: as already noticed in the 1D test-cases, the high-order CCSL-CS quickly reaches its theoretical order of convergence.

\subsection*{3D-3V Boltzmann's equation for neutrals}

As a `real world' application, a 3D kinetic simulation of neutral gas expansion is presented.
Kinetic in this context means we are describing the system at a level where we obtain the distribution of the particles with respect to independent variables including both space and velocity: we solve for the scalar field $f\Par{t, \vect{x}, \vect{v}}$ $[m^{-6}s^3]$, which is the particle number density in phase space.
The kinetic description is necessary because the hypothesis of local thermal equilibrium is not applicable, and hence the form of $f\Par{t,\vect{x},\vect{v}}$ for any point in space cannot be obtained from local integral properties of the fluid, like density, mean velocity and temperature.
Local thermal equilibrium is not reached because the mean free path is comparable with the system dimensions; or, as another way to put it, the collision time is comparable with the time needed to traverse the system. 
 
Using the kinetic formalism, the time evolution of a rarefied single species neutral gas is described by a simplified form of Boltzmann's equation,
\begin{equation} \label{eq:test3D3V}
	\frac{\partial f}{\partial t} + \vect{v}\cdot\frac{\partial f}{\partial\vect{x}} =
	\left[\frac{\partial f}{\partial t}\right]_{\text{coll}} ,
\end{equation}
where all force fields are neglected: the domain is small enough to permit neglect of the gravitational force, and the frame of reference is considered to be inertial. Two-body short-range interactions are taken into account by means of the \emph{Boltzmann collision integral} on the right hand side of the equation. As described in Section \ref{sec:CS_RemapRule}, the integral form of \eqref{eq:test3D3V} is solved using a time-splitting procedure, i.e.\ by successively applying a \emph{ballistic operator} and a \emph{collision operator}. Such a mesh-based procedure has important advantages over particle approaches like the Direct Simulation Monte-Carlo \cite{Bird1994}, e.g.\ the absence of statistical noise, a more uniform resolution in velocity space, and reduced memory requirements for a given accuracy level.
As a downside, the application of the Convective Scheme to neutral simulations poses a number of challenges:
\begin{enumerate}
	\item \label{list:NumDiff} reducing the numerical diffusion in space introduced by the ballistic operator,	
	\item \label{list:CollOp} approximating the Boltzmann collision integral with an efficient collision operator,
	\item \label{list:BCs} treating generic boundary conditions in a semi-Lagrangian framework,	
	\item \label{list:Spurious} reducing spurious coupling effects between spatial and velocity meshes (e.g.\ the `ray effect' in the presence of sharp gradients).
\end{enumerate}
Brief outlines of methods for addressing points \ref{list:CollOp} and \ref{list:BCs}, which are strictly related to the velocity mesh set-up, are given in the appendix.\ Point \ref{list:Spurious} has not been thoroughly investigated by the authors yet, and it is a major topic for future research (but see also \cite{Christlieb2004} where particles were allowed to travel within rays of finite angular extent). Instead, the reduction of numerical diffusion (point \ref{list:NumDiff}) is the principal motivation of this work: a 3D version of the third-order CCSL-CS is used as the ballistic operator of the scheme, and the results are compared to the standard first-order version.

The test case that is shown hereafter is part of the hybrid simulation of a Helicon plasma thruster. Argon gas at ambient temperature is injected through a small injector into a slender quartz tube, which is surrounded by a helical RF antenna and magnets, which create an axial static magnetic field. On the other end, the tube is open to vacuum.
Since it is not uncommon for Helicon plasma thrusters to obtain ionization efficiencies higher than 90\%, the density of neutral gas decreases steeply along the axis of the source: the regime is collisional (low Knudsen number) in proximity to the injector, and it is practically non-collisional (high Kn) toward the exit section.

In the following test-cases no plasma is present, and only Ar-Ar hard-sphere elastic collisions are considered. Accordingly, there is no negative source term due to ionization, and hence the gas density along the axis of the cylinder decreases less steeply than it does in the nominal operational regime of the thruster. In order to test the behavior of the scheme for high Knudsen numbers ($\text{Kn}>1$), an additional simulation with reduced injected flow-rate is run.

The simulation domain is a half cylinder (cut on the meridian plane) of diameter $d=20\text{ mm}$ and length $L=80\text{ mm}$, open on one end to vacuum (perfectly absorbing boundary), and closed on the other end by a plane wall with a small orifice (the injector) at the center. All material walls are assumed adiabatic. The spatial mesh is a uniform Cartesian grid with the $z$ coordinate aligned along the symmetry axis of the cylinder. The cell size is $\Delta x = \Delta y = 0.833\text{ mm}$ and $\Delta z = 1.000\text{ mm}$; a total of 17920 spatial cells is used. The curvature of the external wall of the cylinder is approximated with a `staircase' approach. The velocity mesh is obtained by the independent discretization of the magnitude and the direction of the velocity vector: the speed axis $\|\vect{v}\|$ extends from 0 to 1200 m/s, with 15 subdivisions; the unit vector $\vect{v}/\|\vect{v}\|$ can assume 642 different directions, which are obtained by a hierarchical subdivision of the unit sphere (see \ref{app:VelocityMesh} for more details). Accordingly, the whole velocity space is discretized into 9630 cells. The phase-space mesh is obtained by the Cartesian product of the spatial mesh and the velocity mesh, for a total of approximately 173 million phase-space cells. Since two double-precision values of the distribution function are stored for each phase-space cell, the simulation requires about 2.6 GB of RAM.

Different kinds of boundary conditions are employed in the model:
\begin{enumerate}
	\item at the material walls, \emph{adiabatic diffuse reflection} is implemented: the particles that collide with the walls are reflected back isotropically into the domain, but their kinetic energy is unchanged. In order to deal efficaciously with the staircase wall, an integral formulation is employed (see \ref{app:BoundaryConditions});
	\item at the meridian plane, \emph{periodic boundary conditions} are implemented: particles that travel through such a plane undergo a 180$^{\circ}$ rigid-body rotation about the $z$ axis. This method permits us to simulate only half of the cylinder, exploiting the axial symmetry of the geometry. Unfortunately, the velocity mesh employed does not have a 90$^{\circ}$ symmetry, which would have allowed us to simulate only one quarter of the cylinder;
	\item at the open end of the cylinder, \emph{perfect absorption} is implemented: all particles that cross the exit plane are simply absorbed. This approximation is believed to give quite reasonable results, based on geometrical considerations: because of the high Knudsen number, it is quite improbable that a particle leaving from the exit section will undergo a series of collisions that will direct it back into the domain;
	\item at the injector, the flow-rate and the gas properties are imposed, based on the hypothesis of critical (sonic) flow. This integral formulation is equivalent to imposing \emph{Dirichlet boundary conditions} on the distribution function at the inlet.
\end{enumerate}

The distribution function at the inlet section is assumed to be a drifting Maxwellian. As such, its free parameters are the density $n$, the mean velocity $\vect{u}$ and the temperature $T$, which are uniquely determined from the total temperature $T_0=300$ K, flow-rate $\dot{n}=3\cdot10^{18}$ part/s, section area $A\simeq 2.78\text{ mm}^2$, and Mach number $\textit{Ma}=1$.

The collision operator is based on a simplified Bhatnagar-Gross-Krook (BGK) model \cite{BGK1954}, and its efficient and conservative implementation is described briefly in \ref{app:CollisionOperator}.

\begin{figure}[!ht]
	\includegraphics[width=\textwidth]{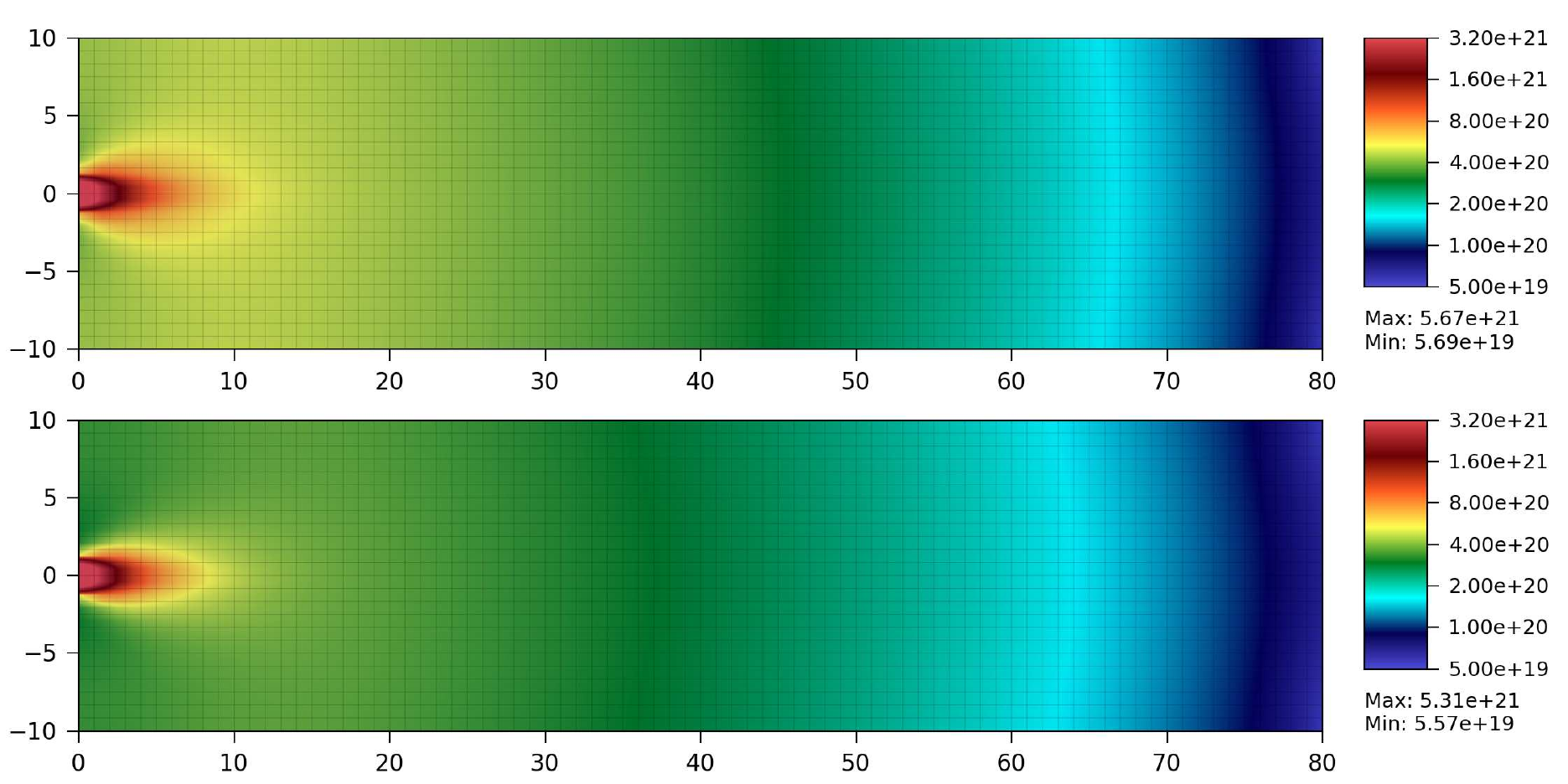}
	\caption{Steady state density distribution for the low-Kn test case: comparison between the 1$^{st}$ order result (top), and the 3$^{rd}$ order result (bottom). Axis scales are in [mm], density is in [m$^{-3}$]. Model grid underlies the plot.}
	\label{fig:dens_lowKn_1st_vs_3rd}
\end{figure}

\begin{figure}[!ht]
	\includegraphics[width=\textwidth]{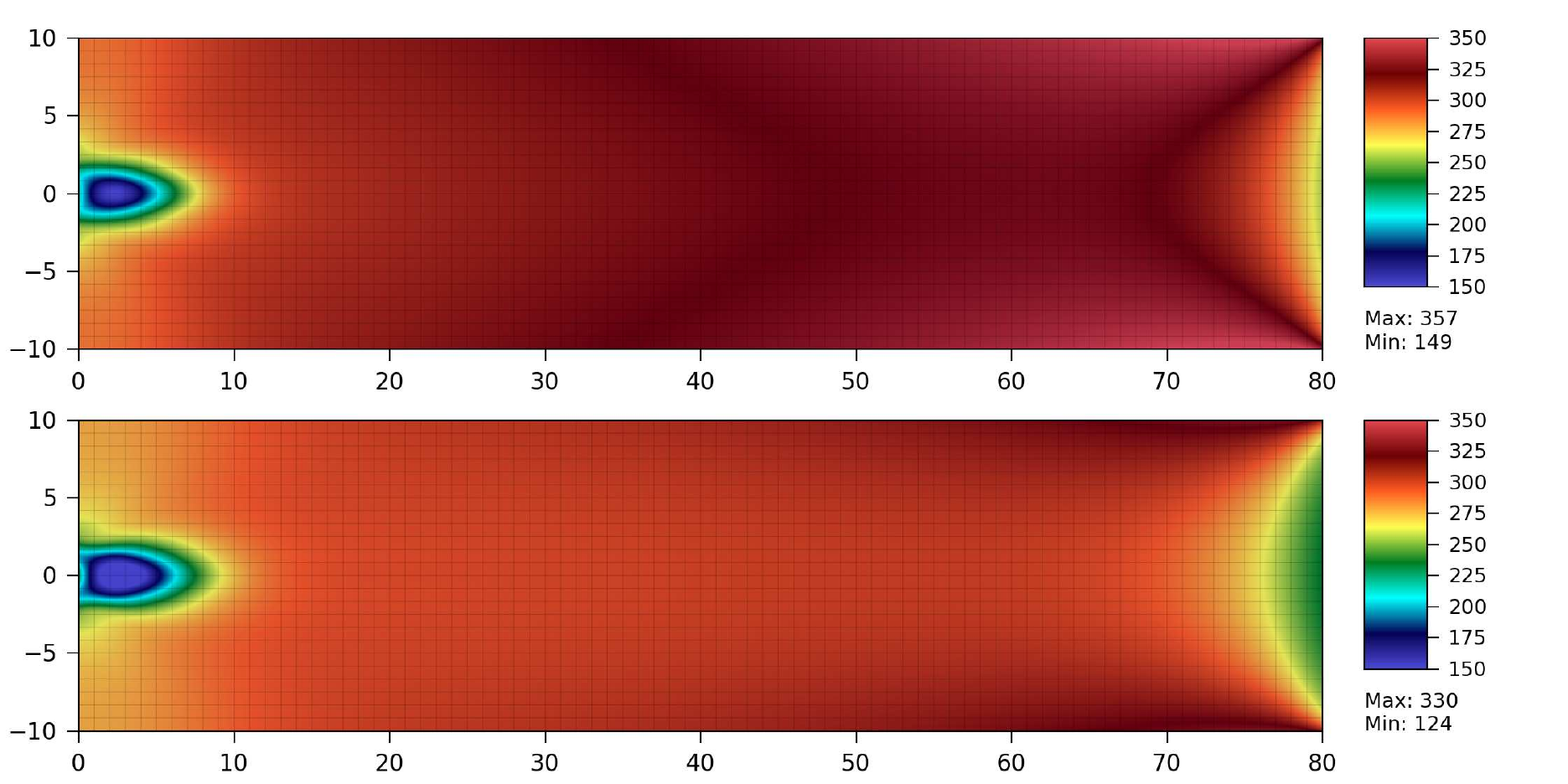}
	\caption{Steady state temperature distribution for the low-Kn test case: comparison between the 1$^{st}$ order result (top), and the 3$^{rd}$ order result (bottom). Axis scales are in [mm], temperature is in [K]. Model grid underlies the plot.}
	\label{fig:temp_lowKn_1st_vs_3rd}
\end{figure}

\begin{figure}[!ht]
	\includegraphics[width=\textwidth]{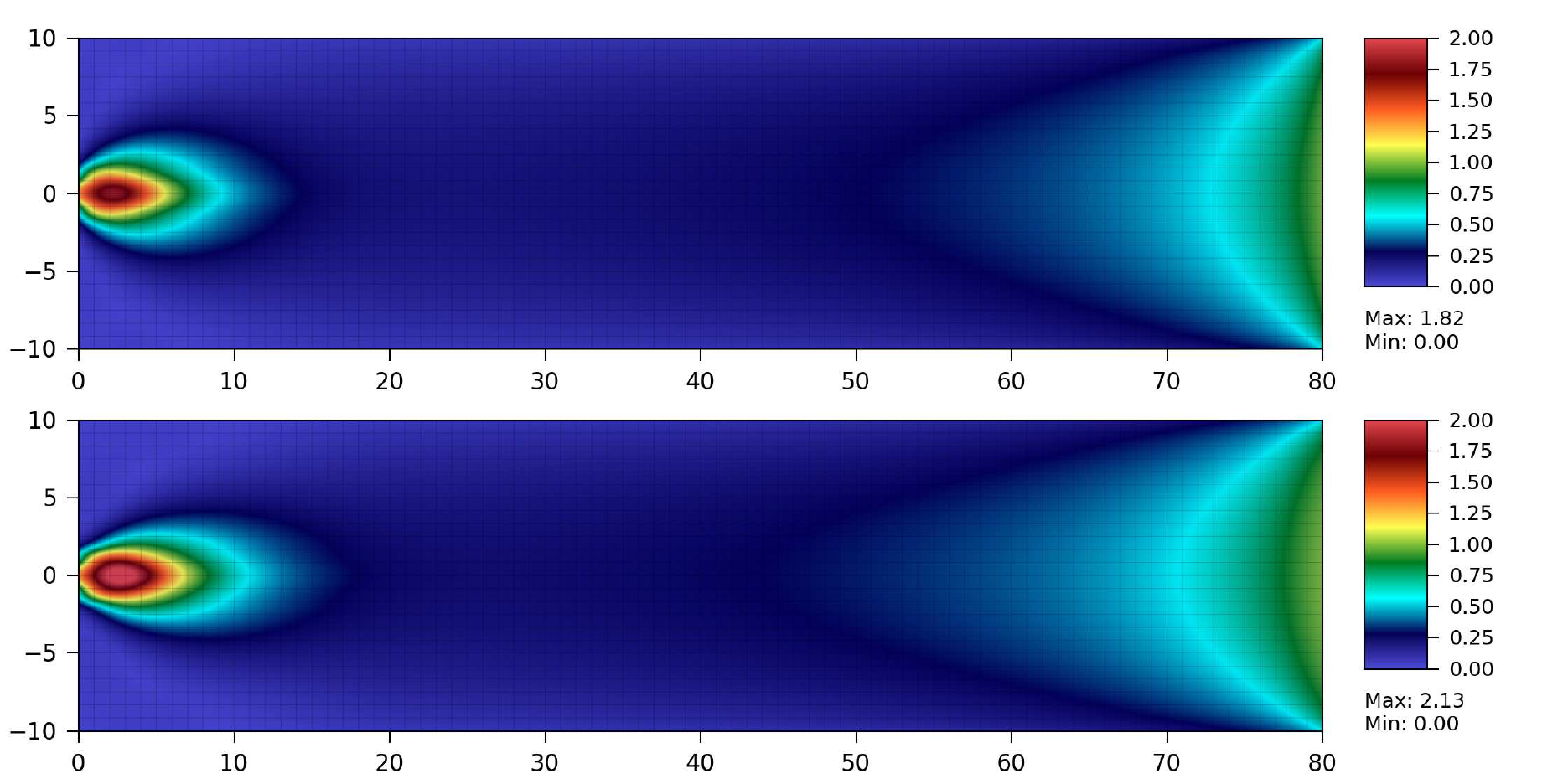}
	\caption{Steady state Mach number distribution for the low-Kn test case: comparison between the 1$^{st}$ order result (top), and the 3$^{rd}$ order result (bottom). Axis scales are in [mm], Ma is dimensionless. Model grid underlies the plot.}
	\label{fig:Mach_lowKn_1st_vs_3rd}
\end{figure}

\begin{figure}[!ht]
	\includegraphics[width=\textwidth]{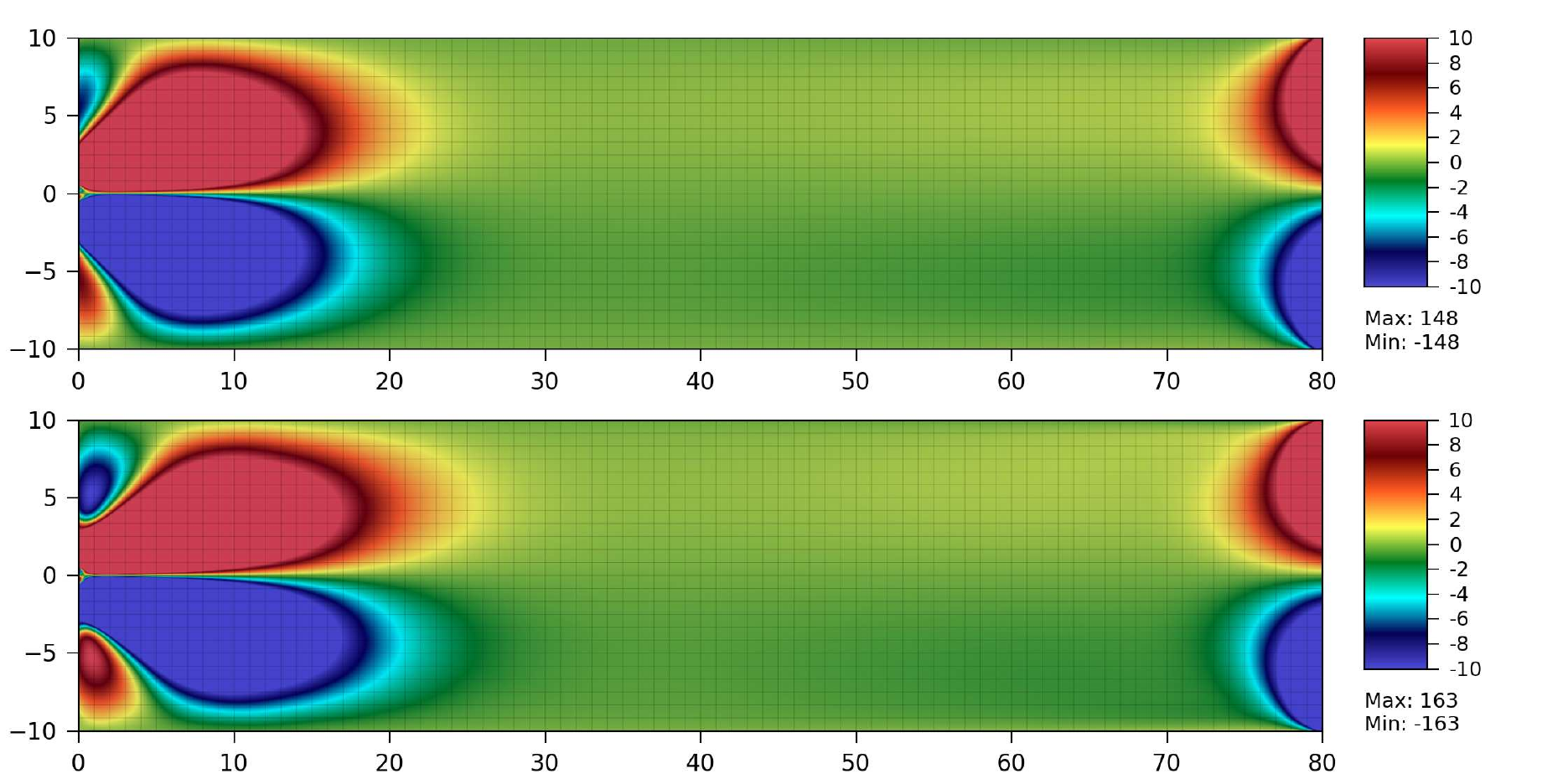}
	\caption{Steady state distribution of the y component of the velocity vector for the low-Kn test case: comparison between the 1$^{st}$ order result (top), and the 3$^{rd}$ order result (bottom). Axis scales are in [mm], velocity is in [m/s]. Model grid underlies the plot.}
	\label{fig:Vy_lowKn_1st_vs_3rd}
\end{figure}

\begin{figure}[!ht]
	\includegraphics[width=\textwidth]{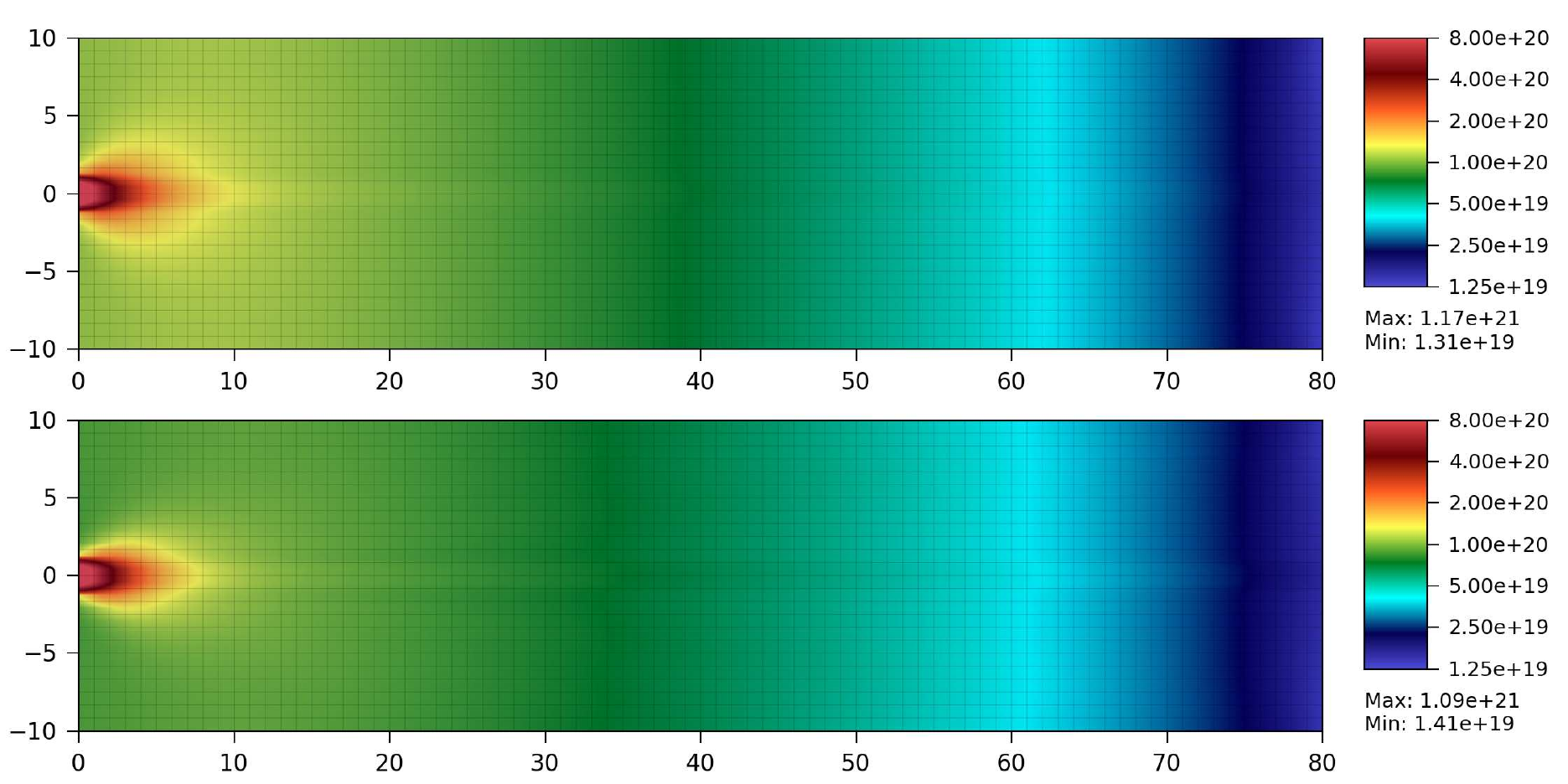}
	\caption{Steady state density distribution for the high-Kn test case: comparison between the 1$^{st}$ order result (top), and the 3$^{rd}$ order result (bottom). Axis scales are in [mm], density is in [m$^{-3}$]. Model grid underlies the plot.}
	\label{fig:dens_highKn_1st_vs_3rd}
\end{figure}

\begin{figure}[!ht]
	\includegraphics[width=\textwidth]{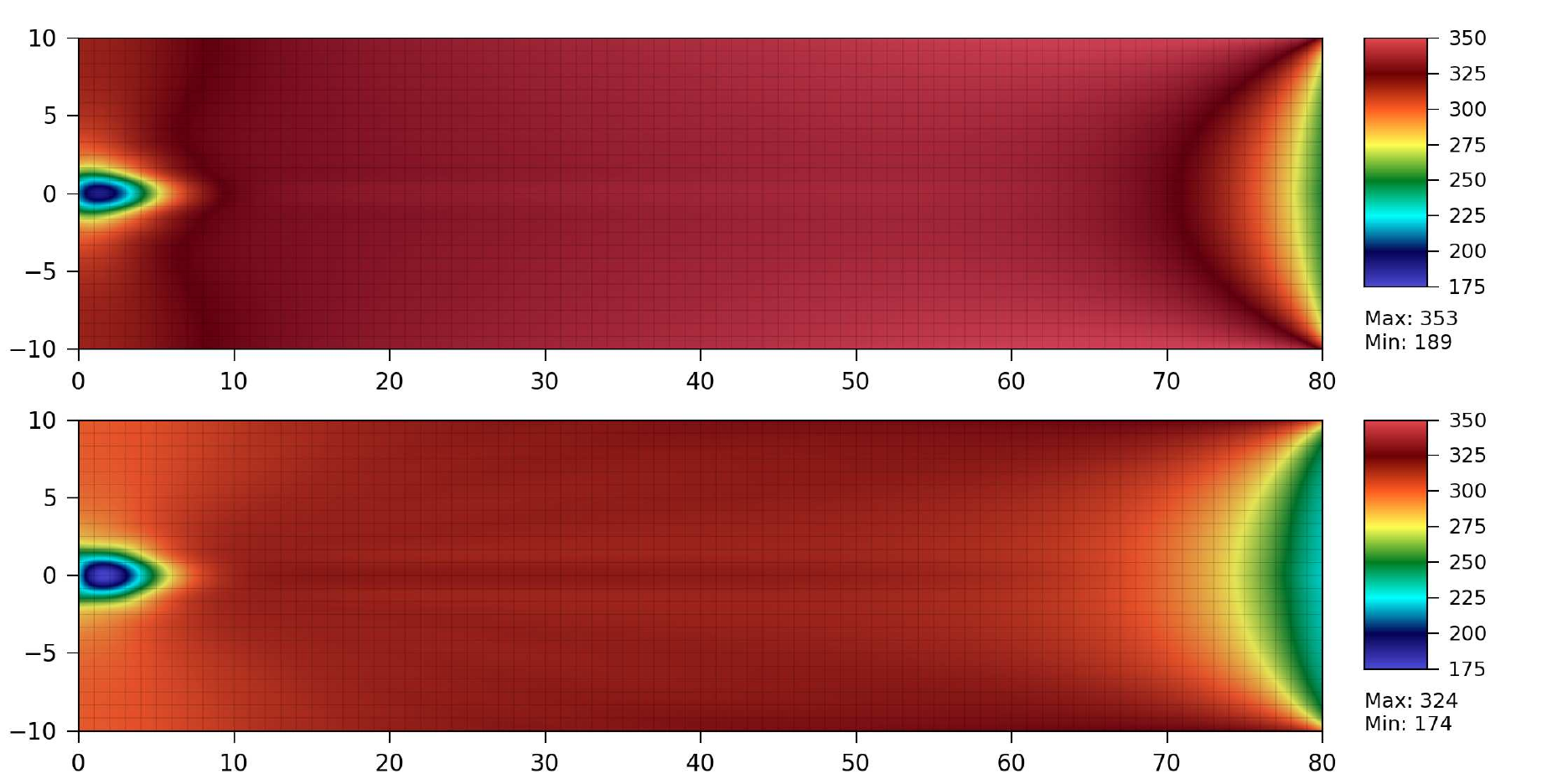}
	\caption{Steady state temperature distribution for the high-Kn test case: comparison between the 1$^{st}$ order result (top), and the 3$^{rd}$ order result (bottom). Axis scales are in [mm], temperature is in [K]. Model grid underlies the plot.}
	\label{fig:temp_highKn_1st_vs_3rd}
\end{figure}

\begin{figure}[!ht]
	\includegraphics[width=\textwidth]{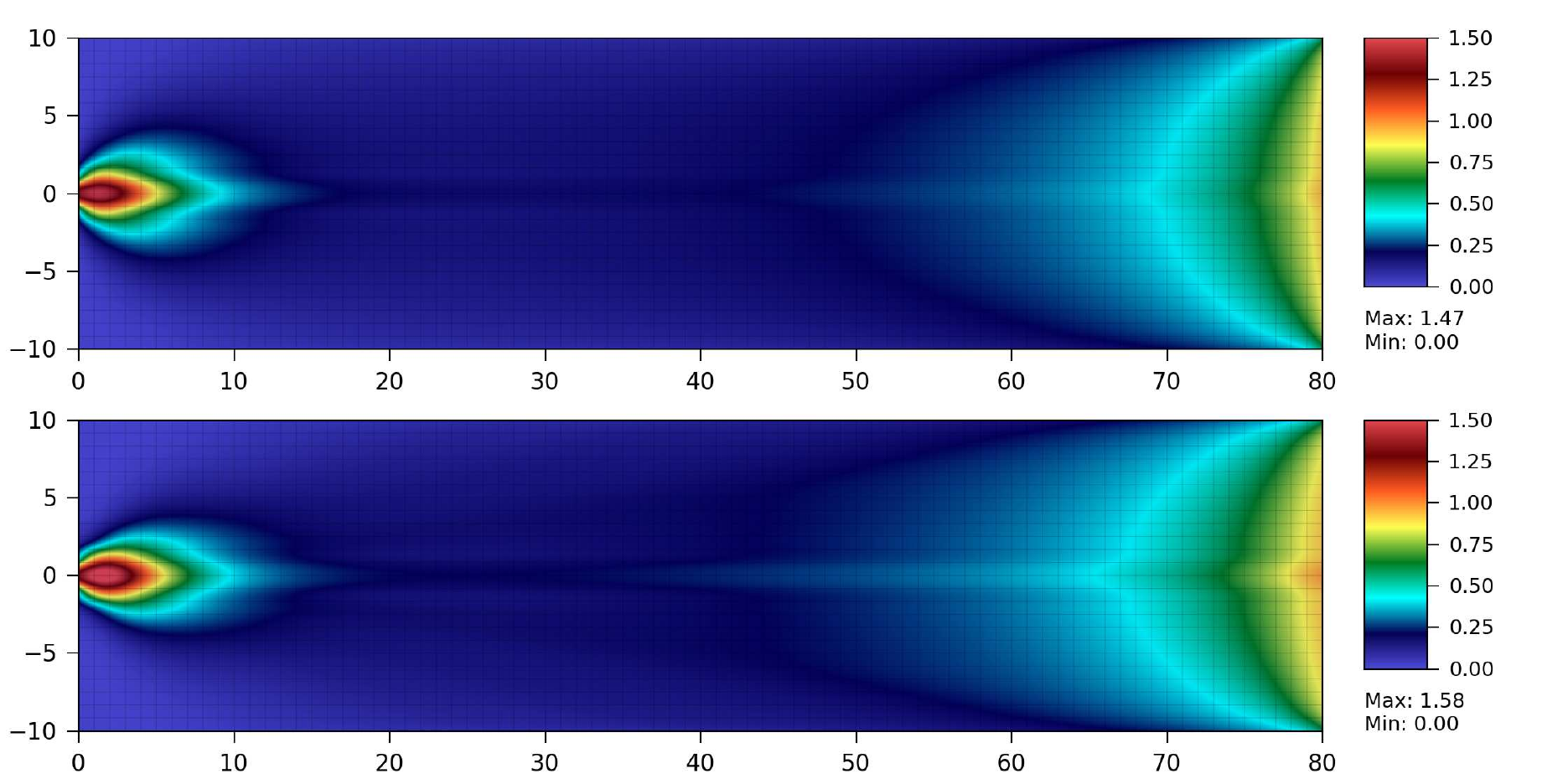}
	\caption{Steady state Mach number distribution for the high-Kn test case: comparison between the 1$^{st}$ order result (top), and the 3$^{rd}$ order result (bottom). Axis scales are in [mm], Ma is dimensionless. Model grid underlies the plot.}
	\label{fig:Mach_highKn_1st_vs_3rd}
\end{figure}

\begin{figure}[!ht]
	\includegraphics[width=\textwidth]{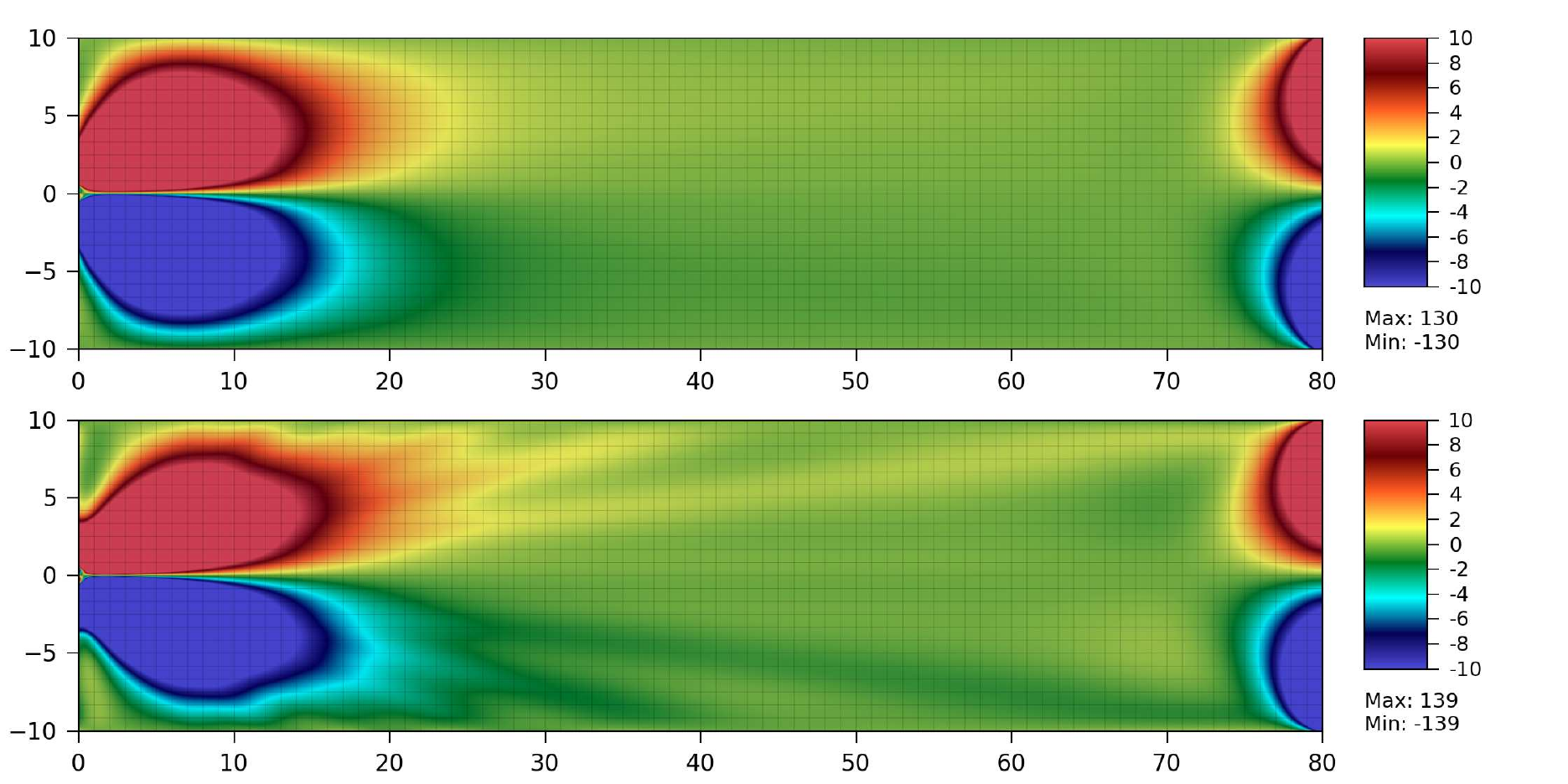}
	\caption{Steady state distribution of the y component of the velocity vector for the high-Kn test case: comparison between the 1$^{st}$ order result (top), and the 3$^{rd}$ order result (bottom). Axis scales are in [mm], velocity is in [m/s]. Model grid underlies the plot.}
	\label{fig:Vy_highKn_1st_vs_3rd}
\end{figure}

Figs.\ \ref{fig:dens_lowKn_1st_vs_3rd}-\ref{fig:Vy_lowKn_1st_vs_3rd} show the steady-state results for the simulations that use the nominal inlet flow-rate $\dot{n}=3\cdot10^{18}$ part/s: due to the low values of the Knudsen number realized in the domain (ranging from $\approx 0.02$ in front of the injector to $\approx 2$ at the exit section), such a situation is labeled as a `low-Kn' test case.
Figs.\ \ref{fig:dens_highKn_1st_vs_3rd}-\ref{fig:Vy_highKn_1st_vs_3rd} show the steady-state results for the simulations that use a reduced inlet flow-rate $\dot{n}=1\cdot10^{18}$ part/s: due to the reduced density in the domain, the Knudsen number ranges from $\approx 0.1$ to $\approx 7$ (`high-Kn' test case).
Two different simulations are run for each test case: one uses the standard first-order ballistic operator, and the other uses the new third-order algorithm. The results are compared in each figure and, as expected, the third order algorithm gives the sharpest results.

Most noticeable are the differences in the region close to the injector, where the high density gradients enhance the effects of numerical diffusion: in both regimes most of the density drop between the injector and the exit section is concentrated just in front of the injector, and this effect is better captured by the third order algorithm (see Figs.\ \ref{fig:dens_lowKn_1st_vs_3rd} and \ref{fig:dens_highKn_1st_vs_3rd}).
As a result of the higher average density obtained in both regimes, the first-order scheme appears to overestimate the temperature by 10-20 K over the whole domain (Figs.\ \ref{fig:temp_lowKn_1st_vs_3rd} and \ref{fig:temp_highKn_1st_vs_3rd}).
At the same time, the first-order scheme appreciably underestimates the peak value of the Mach number in front of the injector (Figs.\ \ref{fig:Mach_lowKn_1st_vs_3rd} and \ref{fig:Mach_highKn_1st_vs_3rd}).
Moreover, in the low-Kn regime, an annular recirculation region appears just next to the injector, and such a phenomenon is much better resolved by the third order scheme (see Figs.\ \ref{fig:Vy_lowKn_1st_vs_3rd}).

In the high-Kn test case the recirculation region vanishes as expected (Fig.\ \ref{fig:Vy_highKn_1st_vs_3rd}), and the low numerical diffusion of the high order scheme makes it possible to observe the spurious `ray-effect' due to the combination of highly divergent flow field, Dirac-delta velocity mesh, and low collisionality.
Such a mesh-based artifact is typical of the so-called `discrete-velocity' models \cite{Broadwell1964} often used in rarefied gas-dynamics: since the velocity can only assume a discrete number of directions, particles which should propagate isotropically from a certain spatial cell can only propagate along the allowed `rays'.
Generally this effect is mitigated by the contribution from the surrounding cells, because the particles coming from those cells tend to `fill-in' the gaps between the previous rays; in fact, there is no ray-effect if the density is uniform.
On the other hand, when the density gradient is large, and specifically when the mean velocity field is highly divergent, the contribution from the surrounding cells may be insufficient and the ray-effect becomes apparent.

When the low-order CCSL-CS is employed (top of Fig.\ \ref{fig:Vy_highKn_1st_vs_3rd}), its high numerical diffusion makes the rays `smear out' in space, so that the gaps between the rays are filled in and the solution looks deceivingly smooth, even for a highly divergent mean velocity field.
As explained in Section \ref{sec:NumericalErrorsAndMotivation}, such numerical diffusion is strongly anisotropic and it depends non-linearly on the speed and direction of the velocity vector; as a consequence, the numerical results depart from the exact solution in a quite unpredictable way.

When the high-order version of the CCSL-CS is employed instead (bottom of Fig.\ \ref{fig:Vy_highKn_1st_vs_3rd}), with very little numerical diffusion in space, only collisions can mitigate the ray-effect: some of the particles change speed and direction by effect of collisions, and in the next time step they end up filling in the gaps between the rays.
Since collisions act on a length scale equal to the mean free path $\lambda$, and rays are only visible when they are separated by more than $\Delta x$, a simple relation should be satisfied to avoid the ray-effect,
\begin{equation}\label{eq:NoRayCondition}
	\Delta\theta < \frac{\Delta x}{\lambda} ,
\end{equation}
which holds for $\Delta x \ll \lambda$.
The inequality \eqref{eq:NoRayCondition} relates the maximum angular spacing $\Delta\theta$ between two adjacent discrete velocities to the amount of \emph{physical diffusion in velocity space} due to particle collisions.
In the high-Kn test-case presented, $\lambda$ is very large and the `no-ray' condition \eqref{eq:NoRayCondition} is in fact not satisfied, as memory requirements made it prohibitive for such a run.

While it may be argued that the low-order scheme `performs better' than the high-order one, because it did not show the unphysical ray-effect in the numerical tests presented, our concern is that the low-order scheme is hiding problems intrinsic in the simulation setup.
Nevertheless, when just a qualitative description of the physical system is sought, a low-order solution affected by numerical diffusion is likely to be preferable over a high-order solution affected by the ray-effect.
Our point here is that we have developed a simple high order scheme, the choice of whether to use it or not in a given instance is a separate matter.


%% file: Conclusions.tex
\section{Conclusions} \label{sec:Conclusions}

A high order scheme for solution of `kinetic equations' was derived, which is 4$^{\text{th}}$ order accurate in space.
The scheme is based on the `Convected Scheme' (CS), which is a method of characteristics applied to the evolution of particle density in phase space.
The CS moves initial cells and maps them back onto the phase space mesh, in such a way as to build in local conservation laws (for density, energy etc.) in a form (discussed in detail elsewhere \cite{Hitchon1999}) such that, for instance, the mean energy is conserved for each final spatial cell, for the given initial phase space cell.
A Modified Equation Analysis applied to the CS was used to show that 4$^{\text{th}}$ order accuracy in space may be obtained by simply applying a small correction to the position of the cell which is to be mapped back to the mesh.

Since the remapping phase was left unchanged, the high-order scheme retains the most desirable properties of the basic CS: it is conservative, it preserves positivity, and it is easy to implement.
In particular, positivity preservation was achieved naturally without the need to strictly enforce monotonicity; this is an advantage over those schemes that use high-order reconstructions to achieve higher-order convergence.
Nevertheless, there exist situations where the numerical solution shows unphysical oscillations (e.g.\ in the presence of shocks); two different `monotonicity-preserving' versions of the CCSL-CS were derived for use in such applications.

Numerical results were illustrated in application to 1, 2 and 3D examples. The order of the scheme appears to be confirmed by these examples.
The 3D example consisted of gas flow from a nozzle through a cylindrical tube into vacuum, and this calculation was performed in Cartesian coordinates and the corresponding velocities. Results were shown at different Knudsen's numbers, illustrating apparent changes in the flow patterns as Kn was varied across the `transitional regime'.

Since the multi-dimensional kinetic simulation of a neutral gas flow is the main application of the present scheme, one of the important conclusions to this work is that we should think carefully about the ray-effect and ways to mitigate it, other than by numerical diffusion.
This will be the subject of future work.

%% file: Acknowledgments.tex
\section{Acknowledgments}

One of us (YG) is grateful to the University of Padua's Doctoral School in Sciences, Technologies, and Measures for Space for support of foreign studies.

%% file: Appendix.tex
\appendix

\numberwithin{equation}{section}


\renewcommand{\thesection}{Appendix \Alph{section}}
\renewcommand{\theequation}{\Alph{section}.\arabic{equation}}



\section{Velocity Mesh} \label{app:VelocityMesh}
%

The velocity grid is a double-indexed list of discrete velocity vectors, $\vect{v}_{ij} = \hat{\vect{n}}_i\,v_j$, where $i$ is the index labeling directions, and $j$ is the index labeling speeds. The unit vectors $\hat{\vect{n}}_i\,$ must be chosen carefully in order to maximize the uniformity of their angular spacing; in order to accomplish this, the surface of a unit sphere is first subdivided into 20 identical triangles (spherical icosahedron), then each triangle is refined into 4 smaller triangles by connecting the midpoints of their sides by great circles. This refinement is repeated recursively, until the required angular resolution is obtained: each vertex on the unit sphere identifies a unit vector $\hat{\vect{n}}_i$, with a solid angle $\Delta\Omega_i$ associated with it. The calculation of $\Delta\Omega_i$ requires the creation of the dual mesh of the given triangular mesh: such a dual mesh is composed of hexagons, plus 20 pentagons, and the areas of those spherical polygons numerically coincide with the solid angles required. The speed axis $v$ is discretized in concentric spherical shells of uniform thickness $\Delta v_j$, and we let $v_j$ be the center radius of the shell. The volume of the cell $(i,j)$ in velocity space [m$^3$s$^{-3}$] can be evaluated as
\begin{equation*}
	V_{ij} = \Delta\Omega_i\,\Delta v_j \Par{v_j^2 + \frac{\Delta v_j ^2}{12}}\ .
\end{equation*}

\section{Collision Operator} \label{app:CollisionOperator}

In an `operator split' environment, the Collision Operator evolves the distribution function for one time-step $\Delta t$ according to \eqref{eq:CollisionStep}. The effect of the Boltzmann Collision Integral on the R.H.S.\ is approximated using an integral formulation of the Bhatnagar-Gross-Krook (BGK) model \cite{BGK1954}, which is implemented in two steps, both local in space:
\begin{enumerate}
	\item the number of particles that scatter out of each phase-space cell during the time-step $\Delta t$, due to collisions against the background gas, is:
	\begin{equation*}
		\Delta N_{\text{out}} = N_0 \left[1-\exp\Par{-\nu\,\Delta t}\right]\ ,
	\end{equation*}
where $N_0$ is the number of particles inside the phase-space cell at the beginning of the time-step, and $\nu\Par{\vect{v}_c}$ is the total collision frequency;
	\item all the particles that have undergone a collision in the same spatial cell are put into a drift-Maxwellian distribution $\mathcal{M}\Par{\vect{v}}$, which gives the number of particles that scatter into each phase-space cell:
	\begin{equation} \label{eq:RemapCollided}
		\Delta N_{\text{in}} = \mathcal{M}\Par{\vect{v}_c}\cdot V_c\ ,
	\end{equation}
	where $V_c$ is the volume of the cell in phase-space [$m^6s^{-3}$].
\end{enumerate}
Both the aforementioned steps give rise to numerical problems. As a first issue, the formula for $\nu\Par{\vect{v}_c}$ is a three-dimensional integral over velocity-space:
\begin{equation} \label{eq:totCollFreq}
	\nu\Par{\vect{v}_c} = \int_{\mathbb{R}^3} \|\vect{v}_c-\vect{v}\|\ \sigma\Par{\|\vect{v}_c-\vect{v}\|}\ f\Par{\vect{v}}\ d^3\vect{v} \ ,
\end{equation}
where $\sigma\Par{\|\vect{v}_c-\vect{v}\|}$ is the cross-section [$m^2$] for elastic collisions between like particles. In the context of the BGK approximation, \eqref{eq:totCollFreq} is a waste of computational effort. In order to approximate the integral on the RHS,  an average value $\langle\sigma\rangle$ is used (justified because the cross section for elastic collisions of non-excited Ar does not change appreciably in the range of velocities of interest). Moreover, the real distribution $f\Par{\vect{v}}$ is approximated by a suitable analytic distribution function $f^*\Par{\vect{v}}$:
\begin{equation} \label{eq:ApproxDistribution}
	f\Par{\vect{v}} \approx f^*\Par{\vect{v}} = 
	\frac{n}{4\pi\,v_T^2}\ \delta_D\big( \|\vect{v}-\vect{u}\|-v_T\big)\ ,
\end{equation}
where $n$, $\vect{u}$ and $v_T$ are the density, mean velocity and thermal velocity of the original $f(\vect{v})$, respectively, and $\delta_D$ is the Dirac delta function. Substituting \eqref{eq:ApproxDistribution} into \eqref{eq:totCollFreq}, introducing the random velocity $\vect{v}' = \vect{v}-\vect{u}$, and using the identity $d^3\vect{v}' = (v')^2 d\Omega\ dv'$, one gets:
\begin{equation} \label{eq:totCollFreq_approx}
	\nu\Par{\vect{v}_c} \approx \frac{n\,\langle\sigma\rangle}{4\pi}
	\oint_{\mathbb{S}} \sqrt{(v'_c)^2 + v_T^2-2\,v'_c\,v_T\,\cos\theta}\ d\Omega\ ,
\end{equation}
which is solvable using $d\Omega = \cos\theta\, d\theta\, d\phi$. Consistently with the approximations already introduced, \eqref{eq:totCollFreq_approx} can be further simplified by neglecting the $\cos{\theta}$ dependence inside the square root, leading to a very easily manageable formula for the total collision frequency:
\begin{equation*}
	\nu\Par{\vect{v}_c} \approx n\,\langle\sigma\rangle \sqrt{(v'_0)^2 + v_T^2}\ .
\end{equation*}

A second issue is that \eqref{eq:RemapCollided} does not ensure mass, momentum and energy conservation on a finite grid: to satisfy such a requirement, one should compute \emph{exactly}
\begin{equation*}
	\Delta N_{\text{in}} = \int_{c}\mathcal{M}\Par{\vect{v}}\,d^3\vect{x}\,d^3\vect{v}
	= \langle\mathcal{M}\rangle_c V_c \neq \mathcal{M}\Par{\vect{v}_c} V_c\ ,
\end{equation*}
but this is extremely difficult, due to the complicated shape of the domain in velocity space. As an efficient alternative to computing the exact integral of the exact Nominal Maxwellian (NM) distribution function $\mathcal{M}\Par{\vect{v}}$, one can still use $\eqref{eq:RemapCollided}$, but substituting the NM with what we refer to as a ``Discrete Conservation-Corrected Maxwellian'' (DCCM), $\mathcal{M}^*\Par{\vect{v}}$. The DCCM is constructed by applying small corrections to the NM that ensure explicitly all the aforementioned conservation rules on the grid. Among other possible choices, we write $\Delta f(\vect{v})$ as a first-order correction to the original Maxwellian: each value of the sampled Nominal Maxwellian is written as a function of the parameters to be conserved,
\begin{equation*}
	f_\alpha\Par{n,\vect{u},\eps'} =  n \Par{\frac{3}{4\pi\eps'}}^{\frac{3}{2}} \exp \left[-\frac{\|\vect{v}_\alpha-\vect{u} \|^2}{4\eps'/3}\right]\ ,
\end{equation*}
and $\Delta f_\alpha$ is the first order term of the Taylor series of $f_\alpha$,
\begin{equation*}
	\Delta f_\alpha =  \frac{\partial f_\alpha}{\partial n} \Delta n + \frac{\partial f_\alpha}{\partial u_x} \Delta u_x +
				\frac{\partial f_\alpha}{\partial u_y} \Delta u_y + \frac{\partial f_\alpha}{\partial u_z} \Delta u_z + \frac{\partial f_\alpha}{\partial \eps'} \Delta \eps'\ .
\end{equation*}
What we have obtained is a 5-term expansion in the basis functions $\frac{\partial f_\alpha}{\partial (\cdot)}$, with unknown coefficients $[\Delta n, \Delta u_x, \Delta u_y, \Delta u_z, \Delta \eps']$. These 5 coefficients are uniquely determined by requiring the combination of all the $\Delta f_\alpha\,$s to satisfy the 5 scalar conservation equations.

\section{Boundary Conditions} \label{app:BoundaryConditions}

For a semi-Lagrangian scheme such as the Convected Scheme, boundary conditions (BCs) are applied to the \emph{characteristics} that come off the wall, and no modification is made due to the wall on the characteristics that go towards the wall. As a consequence, BCs must be handled in a way that satisfies causality, and this is a considerable advantage when compared to Eulerian schemes. On the other hand, the outgoing characteristics must be correlated with the incoming characteristics, explicitly for each case, and this may be a non-trivial problem where quantities like mass or energy are to be exactly conserved.

The use of periodic planes enables the simulation of an axially-symmetric geometry using only half (or one quarter) of the computational domain. Implementing periodic planes is straightforward: when a part of a MC crosses the plane, its position and velocity are transformed by a rotation about the symmetry axis. Issues concerning this procedure take place wherever the transformed velocity does not coincide with one of the discrete values of the velocity mesh: in such cases, it is necessary to remap the velocity vector onto two or more vectors, and this fact introduces numerical diffusion in velocity space (although only in direction, and not in energy). Another possible side-effect is the late-time creation of non-physical density bumps in the region close to the axis, which take place essentially because of the asymmetries in the particle fluxes. The angular mesh used in the current version of the code is symmetric with respect to each of the three principal planes, and this property permits one to simulate only one half of the cylinder without incurring the above drawbacks.

Since in the CS no random numbers are generated, diffuse adiabatic reflection is implemented in an \emph{integral} fashion, which permits us to model any geometry of the wall (a staircase approach is used here) by using a thick layer of ghost-cells:
\begin{enumerate}
	\item the MCs coming from the domain pass seamlessly through the boundary wall, and they remap onto the ghost-cells by applying the same \emph{volume-rule} that is used inside the domain;
	\item the ghost-cells gather the incoming particles on an energy basis, so that the information about the angular direction is lost, but mass and kinetic energy can be exactly conserved;
	\item for every ghost cell and for each speed, a list of the spatial cells in the domain which are `visible' from that cell is constructed. Equal densities are launched into equal solid angles, for those cells in the list. This rule guarantees that an initial uniform isotropic distribution retains those properties after bouncing off the wall.
\end{enumerate}
For any finite mesh-size, the implementation above guarantees that mass and total kinetic energy of the colliding particles are exactly conserved, while their average parallel momentum is transferred to the wall. Moreover, such an algorithm isotropizes the distribution function without introducing any diffusion in energy space, and it has the overall effect of converting directional kinetic energy into random kinetic energy: as a consequence of the friction against the wall (no-slip condition), the temperature of the gas locally increases in an irreversible thermodynamic process. 


The injector effectively imposes Dirichlet BCs on the distribution function. In a semi-Lagrangian framework, this must be implemented with particular care to ensure mass and energy conservation. The solution employed is to load the ghost cells immediately behind the inlet section with a drift-Maxwellian distribution function (having density $n$, mean velocity $\|\vect{u}\|$ and temperature $T$ imposed a priori), and to launch the particles into the domain using the usual ballistic mover. This procedure does not guarantee that the mass flux per unit area [m$^{-2}$s$^{-1}$] equals the expected value $n\|\vect{u}\|$; in fact, the \emph{net mass flux} through any given plane surface $\mathcal{S}_\vect{k}$ with normal $\vect{k}$ is the sum of a positive contribution $\Gamma^+$ due to particles that cross the plane in the direction of $\vect{k}$ (i.e.\ $\vect{v}\cdot\vect{k}>0$), plus a negative contribution $\Gamma^-$ due to particles that cross the plane in the opposite direction ($\vect{v}\cdot\vect{k}<0$). If the same distribution function is loaded on both sides of $\mathcal{S}_\vect{k}$, then $\Gamma^+ + \Gamma^- = n\|\vect{u}\|$ as expected; but in our model the distribution function is loaded only on one side of $\mathcal{S}_\vect{k}$, and $\Gamma^-$ is not imposed: noticing that in a sonic condition we have $\|\Gamma^+\| \gg \|\Gamma^-\|$, a satisfactory solution is to scale the density of the imposed distribution by a correction factor close to one, and not to let particles exit from the injector, so that $\Gamma^+ = n\|\vect{u}\|$ and $\Gamma^- = 0$.

We stress here that, due to the uniform Cartesian spatial mesh employed in the simulations, the injector is modeled in a very simplified way, using just a few cells. (A detailed analysis of the region close to the injector would require a high local resolution, using an unstructured grid as in \cite{Fixel2007}, or an adaptive multi-grid approach).